\documentclass[11pt,conference,a4paper,twoside,onecolumn]{IEEEtran}
\usepackage{epsfig}
\usepackage{amsfonts}
\usepackage{multirow}

\newtheorem{remarks}{Remarks}[section] 
\newtheorem{theorem}{Theorem}
\newtheorem{lemma}{Lemma}
\newtheorem{corollary}{Corollary}
\newcommand{\qed}{\hfill \rule{2.5mm}{2.5mm}}
\newcommand{\prob}[1]{\mathsf{Pr}\left(#1\right)}

\begin{document}

\title{Distributed Construction of \\ the Critical Geometric Graph \\
  in Dense Wireless Sensor Networks}

\author{\IEEEauthorblockN{Srivathsa Acharya\IEEEauthorrefmark{1},
    Anurag Kumar\IEEEauthorrefmark{1}, Vijay
    Dewangan\IEEEauthorrefmark{1}, Navneet
    Sankara\IEEEauthorrefmark{2}, \\ Malati Hegde\IEEEauthorrefmark{1},
    and S.~V.~R.~Anand\IEEEauthorrefmark{1}}
  \IEEEauthorblockA{\IEEEauthorrefmark{1}Department of Electrical
    Communication Engineering, Indian Institute of Science,
    Bangalore, India \\ Email: srivatsa.acharya@gmail.com,
    anurag@ece.iisc.ernet.in, vijay.dewangan@gmail.com}
  \IEEEauthorblockA{\IEEEauthorrefmark{2}Department of Electrical
    Engineering, Indian Institute of Technology,
    Madras, India; Email: navsan@gmail.com}
}

\maketitle
\vspace{-1mm}

\begin{abstract}
  Wireless sensor networks are often modeled in terms of a dense
  deployment of smart sensor nodes in a two-dimensional region. Given
  a node deployment, the \emph{critical geometric graph (CGG)} over
  these locations (i.e., the connected \emph{geometric graph (GG)}
  with the smallest radius) is a useful structure since it provides
  the most accurate proportionality between hop-count and Euclidean distance.
  Hence, it can be used for GPS-free node localisation as well as minimum
  distance packet forwarding. It is also known to be
  asymptotically optimal for network transport capacity and power
  efficiency. In this context, we propose DISCRIT, a distributed
  and asynchronous algorithm for obtaining an approximation of the CGG
  on the node locations. The algorithm does not require the knowledge
  of node locations or internode distances, nor does it require
  pair-wise RSSI (Received Signal Strength Indication) measurements to be made.
  Instead, the algorithm makes use of successful Hello receipt counts
  (obtained during a Hello-protocol-based neighbour discovery process)
  as edge weights, along with a simple distributed min-max computation
  algorithm. 


  In this paper, we first provide the theory for justifying the use of
  the above edge weights. Then we provide extensive simulation results
  to demonstrate the efficacy of DISCRIT in obtaining an approximation
  of the CGG. Finally, we show how the CGG obtained from DISCRIT
  performs when used in certain network self-organisation algorithms.

\end{abstract}
\vspace{-2mm}
\section{Introduction} 

\label{sec:introduction}

In a wireless sensor network, the smart sensor nodes (often called
\emph{motes}) are embedded in some space, region or structure in order
to make measurements and to draw inferences. In this paper, we are
concerned with situations in which there is a dense deployment of
nodes over a 2-dimensional region. We propose and study an algorithm,
called DISCRIT, for the distributed construction of an approximation
to the \emph{critical geometric graph (CGG)} over the set of node
locations.  

\noindent
\textbf{Notation:} We denote the region of deployment by ${\cal A} \in
\mathbf{R}^2$.  Given a set of $n$ nodes $N= \{1,2,\dots,n\}$ and the
node location vector $\mathbf{V} = \left(V_1,V_2,\dots,V_n \right) \in
{\cal A}^n$, the \emph{geometric graph (GG)} of radius $r$, denoted by
${\cal G}(\mathbf{V},r)$, is obtained by joining any two nodes within
a distance of $r$. That is, the edge set $E$ of ${\cal
  G}(\mathbf{V},r)$ is given by $E= \{(i,j) \in N^2: d_{i,j} \leq
r\}$, where $d_{i,j}$ represents the Euclidean distance between nodes
$i$ and $j$. A sufficiently small $r$ can cause the resulting GG to be
disconnected.  Thus, the smallest $r$ on the given node locations at
which the corresponding GG becomes connected is called the
\emph{critical radius}, denoted as $r_{crit}(\mathbf{V})$. We will
call the corresponding GG, ${\cal G}
(\mathbf{V},r_{crit}(\mathbf{V}))$, the critical geometric graph (CGG)
on $\mathbf{V}$. When the node locations are implicit, we denote the
critical radius as $r_{crit}$ and the CGG as ${\cal G}_{crit}$.

\noindent 
\textbf{Motivation for ${\cal G}_{crit}$:} The following are some
reasons why it would be useful to develop a distributed algorithm for
obtaining the CGG on a given set of node locations.

\begin{enumerate}

\item In most applications of wireless sensor networks, it is
  important for the nodes to be aware of their own locations and 
  distances to other nodes (in particular the network's fusion 
  centers, or base stations).  While accurate and low-power Global
  Positioning System (GPS) support is becoming available on motes, such an
  approach may not only be expensive and power hungry, but also, not
  workable for all applications, for example in large indoor areas or in
  mines.  One approach for GPS-free distance estimation and node
  localisation is to overlay a geometric graph on the node locations.
  It can then be shown that the number of edges along the shortest
  path between an \emph{anchor point} on the plane and a node is
  roughly proportional to the Euclidean distance between the point and
  the node, \emph{the proportionality factor being the radius of the
    GG} (see \cite{wsn.swaprava-kumar08hd-ed-proportionality}).  The
  usual approach for obtaining a GG on the set of node locations
  (e.g., \cite{wsn.niculescu-nath01aps} and
  \cite{wsn.nagpal-etal03localization}) yields a radius equal to the
  radio-range of the nodes.  Since the CGG is the connected geometric
  graph (on the given node locations) with the smallest radius, it
  provides the ``finest scale'' with which to measure distances using
  hop-counts on a GG on the nodes.

\item Further, we observe that, since hop-counts on the CGG provide a
  measure of Euclidean distance, these hop-counts can also be used in
  topology-free routing. For example, a node that has a smaller hop-count 
  (on the CGG) to the base-station is also likely to be 
  closer to the base-station, and hence can serve as a forwarding node in a
  topology-free routing algorithm (see \cite{wsn.ye-etal05grab}).

\item We also recall that, under the setup used by Gupta and Kumar
  \cite{wsn.gupta-kumar00capacity}, ${\cal G}_{crit}$ turns out to be
  an asymptotically optimal topology for maximising the transport
  capacity of the network, and minimizing the network power
  consumption. 

\end{enumerate}

\noindent
\emph{Remark:} Note that while Point~3) above suggests the use of
${\cal G}_{crit}$ itself as the communication topology, in Points~1)
and 2) ${\cal G}_{crit}$ is used only as a means for obtaining
distance measurements on the region on which the nodes are deployed.

\noindent
\textbf{System assumptions:} To obtain a distributed algorithm to
approximate ${\cal G}_{crit}$, we make the following assumptions. We
consider a \emph{dense} node deployment, where the nodes are deployed
in excess of the minimum requirement for connectivity and sensing
coverage. For the purpose of developing the theory, we consider the
\emph{uniform i.\ i.\ d.\ } deployment where each node is placed
randomly uniformly on the region ${\cal A}$, independent of the
placement of the other nodes. However, simulation results are shown
for the \emph{randomised lattice} and \emph{grid} deployments as well.
In a randomised lattice deployment, the region ${\cal A}$ is divided
into $n$ partitions of equal area called cells, and one node is placed
randomly over each cell. The grid is a deterministic deployment where
the nodes are placed in $\sqrt{n}$ rows and $\sqrt{n}$ columns on
${\cal A}$, with the rows and columns equally spaced.

The channel model includes path loss, and fading with additive
Gaussian noise, where the fading process is assumed to be stationary
in space and time with a common marginal distribution. The terrain is
assumed to be flat, and the node transmission is assumed to be
omnidirectional, so that the power radiated in all directions is
equal.

\noindent
\textbf{Contributions:} With the above assumptions, our contributions
are the following.  We develop DISCRIT (DIStributed CRITical geometric
graph algorithm), a distributed algorithm for constructing an
approximation to ${\cal G}_{crit}$. The algorithm is based on a result
due to Penrose \cite{wsn.penrose99k-connectivity} which holds for
uniform i.\ i.\ d.\ deployments, and states that as the number of
nodes $n \to \infty$ the CGG becomes the same as the \emph{farthest
  nearest neighbour geometric graph (FNNGG)}. Given the internode
distances, the FNNGG can easily be constructed by a distributed
max-min computation, thus providing an approximation to the CGG, for
large $n$.  Since we do not know internode distances, we utilise a
technique that provides us with a monotone \emph{decreasing} function
of the internode distances, thus permitting the use of a
\emph{distributed min-max computation} to obtain the FNNGG. Such a
function is obtained by using Hello reception counts obtained during
the Hello-protocol-based neighbour discovery (see
\cite{wsn.karnik-kumar04self-organization}).  The Hello transmissions
can proceed completely asynchronously (e.g., via CSMA broadcasts),
thus not requiring transmission synchronisation, as might be necessary
in an RSSI-based approach. We show theoretically (using our
assumptions, above) that the counts we obtain can serve as surrogates
for the internode distances directly in the special case of isotropic
antenna radiation patterns.  Then we provide extensive simulations
results to support this theory and our overall DISCRIT proposal.

Finally, we provide numerical evaluations of two applications of the
approximate CGG obtained from DISCRIT:

\begin{enumerate}
\item Optimal forwarding hop distance determination, as per the theory
  provided by Ramaiyan et al.\ in
  \cite{wsn.ramaiyan-etal07optimal-transport-capacity}
\item Hop Count Ration based Localisation (HCRL)
  (\cite{wsn.yang-etal07hcrl}).
\end{enumerate}


\noindent
\textbf{Related Literature:} Narayanaswamy et al.\
\cite{wsn.narayanaswamy-etal02compow} provide the COMPOW protocol for
obtaining ${\cal G}_{crit}$. The idea here is to operate all nodes at
the lowest common power level of available discrete power levels while
ensuring connectivity in the network. As the communication range is an
increasing function of the transmission power, the minimum common
power results in the minimum range for connectivity, and thus yields
${\cal G}_{crit}$. However, COMPOW requires distance-vector routing to
be done for each available discrete power level; switching between
power levels requires synchronisation among the nodes. Unlike COMPOW,
the proposed algorithm DISCRIT does not require multiple power levels
and synchronisation between nodes. Also, DISCRIT needs to be run only
once by the nodes to obtain ${\cal G}_{crit}$, unlike COMPOW where
routing has to be done for each power level. The literature related to
using hop-distance as a measure of inter-node distance is discussed
later in Section~\ref{sec:distance_discretisation}. 


\noindent
\textbf{Outline of the Paper:} Section~\ref{sec:algorithm} gives the
algorithm DISCRIT for obtaining an approximation to ${\cal G}_{crit}$
along with the associated theory
(\ref{sec:distance_free_algorithm_iso}).  Section~\ref{sec:simulation}
provides simulation results on the performance of DISCRIT for various
deployments. The CGG based distance discretisation technique and its
justification are provided in
Section~\ref{sec:distance_discretisation}.  Finally, as an application
of distance discretisation using DISCRIT, we provide numerical results
for (i) a self-organisation formulation
(Section~\ref{sec:optimal_self_organisation}), and (ii) an approach
for approximate node localisation (Section~\ref{sec:localisation}).
Section~\ref{sec:conclusion} concludes the paper with future work.

\section{Theory and the Algorithm} \label{sec:algorithm} In this
section, we first arrive at an algorithm for ${\cal G}_{crit}$ by
making use of a result by Penrose
\cite{wsn.penrose99k-connectivity}. The algorithm requires each node
to know the distances to its neighbours. The distance-free distributed
algorithm (DISCRIT) is then obtained by using link-weights obtained
from a Hello-protocol-based neighbour discovery as distance-like
information.
\vspace{-1mm}
\subsection{Degree-1 GG and Penrose's Theorem}
Given node locations $\mathbf{V}$, let $r_1 (\mathbf{V})$ be the
smallest $r$ such that the corresponding GG, ${\cal G}(\mathbf{V},r)$
has no \emph{isolated} node, i.e., ${\cal G}(\mathbf{V},r)$ has the
\emph{degree}~\footnote{The degree of a node in a graph is the number
  of its \emph{adjacent} nodes. The degree of a graph is the minimum
  of its node degrees.}  of at least 1. It can be seen that $r_1
(\mathbf{V})$ is precisely the maximum of the nearest node distances,
i.e., \vspace{-3mm}
\begin{equation} \label{eqn:r_1}
r_1 (\mathbf{V}) = \max_{i \in N} \left \{\min_{j \in N, j \ne i } \{d_{i,j}\}\right\}
\end{equation}
We call the corresponding GG, ${\cal
G}(\mathbf{V},r_1(\mathbf{V}))$ as the \emph{degree-1 GG}, and
denote it by ${\cal G}_1$ and $r_1 (\mathbf{V})$ by $r_1$ when the
node locations are implicit.

Consider a uniform i.\ i.\ d.\ deployment of $n$ nodes. Thus the
random node location vector $\mathbf{V}$ corresponds to the
probability space $({\cal A}^n, {\cal F}^n, {\cal P}^n)$ where ${\cal
  P}$ is the uniform measure on ${\cal A}$, ${\cal P}^n$ is the
corresponding product measure, and ${\cal F}^n$ is the Borel field in
${\cal A}^n$. Theorem~\ref{thm:k-connectivity} gives the relationship
between ${\cal G}_1$ and ${\cal G}_{crit}$ for uniform i.\ i.\ d.\
deployment.

\begin{theorem} [Penrose \cite{wsn.penrose99k-connectivity}] \label{thm:k-connectivity}
Let $\rho_k(\mathbf{V})$ be the minimum $r$ at which $ {\cal
  G}(\mathbf{V},r)$ is \emph{$k$-connected} \footnote{A graph is
  $k$-connected if there exist $k$ \emph{independent} paths between
  any two nodes.}, and $\sigma_k(\mathbf{V})$ be the minimum $r$ at
  which $ {\cal G}(\mathbf{V},r)$ has degree $k$. Then
\[\lim_{n \to \infty} {\cal
  P}^n\{\mathbf{V}: \rho_k(\mathbf{V})=\sigma_k(\mathbf{V})\} = 1 \] 
\end{theorem} $\hfill \qed $

\noindent
Since $\rho_1(\mathbf{V}) = r_{crit}(\mathbf{V})$ and
$\sigma_1(\mathbf{V})=r_1(\mathbf{V})$, we have
\begin{corollary} \label{cor:relation between r_1 and r_crit}
$\lim_{n \to \infty} {\cal
   P}^n\{\mathbf{V}: r_{crit}(\mathbf{V})=r_1(\mathbf{V})\} =1 \hfill \qed$
\end{corollary}

Thus, Corollary~\ref{cor:relation between r_1 and r_crit} indicates
that, for a dense node deployment, ${\cal G}_1$ is identical to ${\cal
  G}_{crit}$ w.h.p.\ \footnote{w.h.p.\ stands for ``with high
  probability,'' i.e., ``with probability $\to$ 1 as $n \to
  \infty$''.}. Note that in general, a graph with no isolated nodes
need not be connected. But the result above implies that, if the graph
is a GG and the node deployment is uniform i.\ i.\ d.\, then just
having no isolated nodes ensures connectivity w.h.p.\ We thus look
for a distributed construction of ${\cal G}_1$, as it is the same as
${\cal G}_{crit}$ w.h.p.

Now, suppose each node $i$ knows the distances $d_{i,j}$ to each of
 its neighbours $j$. Then a node would know its nearest-neighbour
 distance too. As $r_1$ is the maximum nearest-neighbour distance from
 (\ref{eqn:r_1}), a \emph{distributed maximum-finding algorithm} can
 be run by each node to obtain $r_1$. One such distributed
 maximum-finding algorithm is described in Section~\ref{sec:algorithm
 using distances}. Once $r_1$ is known, each node includes all nodes
 within a distance of $r_1$ as its \emph{adjacent nodes}. This results
 in a GG of radius $r_1$, which is ${\cal G}_1$.
\vspace{-1mm}
\subsection{An Algorithm for ${\cal G}_1$ using Distance Information}  \label{sec:algorithm using distances}
Here, every node $i$ maintains a \emph{range} $r(i)$ and an
\emph{adjacent node list} $N(i)$ which get updated as the algorithm
progresses. At any iteration, $N(i)$ is the set of nodes whose
distances are less than range $r(i)$ from node $i$.

\begin{enumerate}
\item \textbf{Initialisation:}
For every node, the range is
 initialised to its nearest neighbour distance, and the adjacent node
 list contains only the nearest neighbour(s) and itself. That is, for
 all $i \in N$,
\[r^{(0)}(i) = \min_{j \in N, j \neq i} \{d_{i,j}\} \qquad \mbox{and}  \qquad N^{(0)}(i)= \{j \in N: d_{i,j} \leq r^{(0)}(i)\}  \vspace{-1mm}\]
Set the iteration index $k=0$

\item \textbf{Current range unicast:}
Every node $i$ informs its
current range $r^{(k)}(i)$ to all its current \emph{adjacent nodes},
i.e., the nodes in $N^{(k)}(i)$.  Therefore, the node $i$ also
receives values of current ranges from some of the nodes which belong to
the set $S^{(k)}(i)~ =~\{j\in N:i \in N^{(k)}(j)\}$.

\item \textbf{Updating the Adjacent Node List:}
Every node $i$ then updates its current range $r^{(k+1)}(i)$ to the
maximum of the ranges it received. The maximum finding includes the
node's current range $r^{(k)}(i)$ also. The adjacent node list
$N^{(k+1)}(i)$ is also updated accordingly as the set of nodes whose
distances from $i$ are within $r^{(k+1)}(i)$.
\[r^{(k+1)}(i)= \max\{r^{(k)}(j): j\in S^{(k)}(i)\} \qquad \mbox{and}\qquad N^{(k+1)}(i) = \{j \in N: d_{ij} \leq r^{(k+1)}(i)\}\]

\item \textbf{Terminating condition:} The algorithm terminates if all
the ranges in an iteration remain unchanged, i.e.,

IF $r^{(k+1)}(i) = r^{(k)}(i)$ for all $i \in N$, \hspace{5mm} Call $r(i)=r^{(k)}(i)$ and $N(i)=N^{(k)}(i)$; \hspace{3mm}\textbf{STOP}\\
ELSE \hspace{5mm}Set $k:=k+1$; \hspace{5mm} go to  Step \textbf{2}.

Note that this is a \emph{centralised} terminating condition. A
\emph{distributed} terminating condition is discussed later.

\item \textbf{Obtaining the topology:} The graph resulting from the
algorithm $G_A$ is obtained by joining each node $i$ to each node in
its final adjacent node list $N(i)$. i.e, ${\cal G}_A =
(\mathbf{V},E_A)$ where $E_A=\{(i,j): i\in N, j\in N(i)\}$.$\hfill \qed$
\end{enumerate}

The convergence of the algorithm output to ${\cal G}_1$ is given in
the following theorem.
\begin{theorem} \label{thm:algorithm}
If ${\cal G}_1$ is connected, then the algorithm converges to ${\cal G}_1$, i.e.,
  ${\cal G}_A = {\cal G}_1$, in at most $D$ iterations, where $D$ is
  the \emph{hop diameter} of ${\cal G}_1$.
\end{theorem}
\begin{IEEEproof}
See Appendix~\ref{appendix:proof} for the proof.
\end{IEEEproof}

From Corollary~\ref{cor:relation between r_1 and r_crit}, ${\cal G}_1$
is indeed connected with high probability, i.e., ${\cal G}_1 = {\cal
  G}_{crit}$ w.h.p. Thus from Theorem \ref{thm:algorithm}, the
algorithm above converges to ${\cal G}_{crit}$ w.h.p.

\begin{remarks}
One can construct deployments where the algorithm fails to give a connected graph. However, these cases happen only when ${\cal G}_1$ is not connected (for if ${\cal G}_1$ is connected then Theorem~\ref{thm:algorithm} holds); the probability of this, as discussed above, is very small for dense networks (probability $\to$ 0 as $n \to \infty$).
\end{remarks}
\begin{remarks}
  Gupta and Kumar \cite{wsn.gupta-kumar98connectivity} have shown that
  $r_{crit}$ scales as $r(n) = \Theta(\sqrt{\frac{\log n}{n}})$ w.p.1
  as $n \to \infty$. Hence the hop diameter of ${\cal G}_{crit}$
  scales as $\Theta(\frac{1}{r(n)}) = \Theta(\sqrt{\frac{n}{\log n}})$
  w.h.p. Thus, the algorithm converges in $\Theta
  \left(\sqrt{\frac{n}{\log n}} \right)$ iterations w.h.p.
\end{remarks}
\begin{remarks}
Note that after updating its range $r^{(k)}(i)$ to
  $r^{(k+1)}(i)$, the node $i$ needs to communicate its new range (i.e., execute Step~\textbf{2}) only
  if it is different from the earlier range. In such a case, there
  will be no communication when the terminating condition in Step~4 is
  met. Hence, a \emph{distributed} terminating condition would be
  that, after informing its current range, each node waits for a
  certain \emph{time-out} period, and locally decides to terminate the
  algorithm if it receives no communication from any node during this
  period.
\end{remarks}

\vspace{-1mm}
\subsubsection{Extension to Monotone Functions of
  Distances} \label{sec:extension of algorithm}
Theorem~\ref{thm:algorithm} can be shown to hold even in the case
where the distances $d_{i,j}$ in the algorithm are replaced by
$f(d_{i,j})$, where $f$ is \textbf{monotone increasing}. Thus, if we
have a \emph{distance-like} information (a monotone function of
distance) known at each node about all its neighbours, then a
\emph{distance-free} distributed algorithm for ${\cal G}_{crit}$ can
be obtained. For this purpose, we consider the Hello-protocol-based
neighbour discovery proposed by Karnik and Kumar
\cite{wsn.karnik-kumar04self-organization}, and use certain \emph{link
  weights} obtained from the neighbour discovery as surrogates for
distances. We assume that the antennas of all the motes  have
isotropic radiation patterns. Such radiation isotropy is a valid
assumption for external ``stick'' antennas.  The resulting
distance-free algorithm, DISCRIT, is described in
Section~\ref{sec:distance_free_algorithm_iso}.  The
Hello-protocol-based neighbour discovery is described next, along with
a discussion of monotonicity of link weights under these
conditions. Under antenna pattern anisotropy (e.g., on-board ``patch''
or ``integrated circuit'' antennas), however, these link weights are
found not to be monotonic with distance. In related work (not reported
here) we have pursued an approach to collate the link weights from
neighbouring nodes in order to obtain a monotonic function of
distance.

\vspace{-1mm}
\subsection{Hello-Protocol-Based Neighbour
  Discovery} \label{sec:neighbour discovery} For the present
discussion, we consider a slotted system. Note that this is \emph{not}
a necessary requirement.  Indeed, the algorithm developed here easily
applies to CSMA/CA scheduling.  We have implemented the algorithm on a
\emph{Qualnet} simulator where IEEE 802.11b CSMA/CA is used, and
\emph{the results reported later in Section~\ref{sec:simulation} are
  with the CSMA/CA}.  In each slot, a node chooses to be either in the
transmit mode with probability $\alpha$, or in receive mode with
probability $1-\alpha$. Whenever in the transmit state, the node
broadcasts a Hello packet where the Hello packet could simply be a
packet containing the node \emph{id}. Let $H_{i,j}$ represent the
fading coefficient from $i$ to $j$, with cumulative distribution
$A(.)$, which is assumed to be identical across all transmit-receive
pairs $(i,j)$. To model if a Hello packet of a transmit node $i$ has
been successfully decoded by a receive node $j$, the \emph{physical
  model} for communication can be used.  That is, transmission from
$i$ to $j$ in a slot is successful if the
signal-to-interference-plus-noise ratio (SINR) of $i$ at $j$ is above
a threshold $\beta$, i.e., if \vspace{-2mm}
\[SINR (i,j) = \frac{H_{i,j} P_t/{d_{i,j}^\eta}}{\sigma^2 + \sum_{k\in N, k\ne
  i}(H_{k,j} P_t/d_{k,j}^\eta) I_k} \ge \beta  \vspace{-1mm}\]
where $P_t$ is the node transmit power at a reference distance, $\eta$ is the path loss
exponent, and $\sigma^2$ is the noise variance. $I_k$ is the indicator function of
the event that the node $k$ is simultaneously transmitting in the same
slot, given that $i$ is transmitting in the slot.

This process of broadcasting Hello packets is carried out for a
large number of slots $t$. Let $C_{i,j}(t)$ be the count of Hellos
of $i$ received at $j$. Let $B_i(t)$ be the number of Hellos broadcast by
$i$ during this period. At the end of $t$ slots, each receive node $j$
calculates $\frac{C_{i,j}(t)}{B_i(t)}$ which is the fraction of Hellos
of $i$ decoded by $j$.

Let $p_{i,j}$ be the probability that $i$'s Hello is successfully
received at $j$, under the broadcast setting described above. Thus
$p_{i,j}$ is the probability that in the slot in which $i$ is
transmitting, $j$ is in receive mode, and the SINR of $i$ at $j$ is
greater than $\beta$, i.e.,
\vspace{-1mm}
\begin{equation} \label{eqn:p_{i,j}}
p_{i,j}= (1-\alpha)\mathsf{Pr}\left(\frac{H_{i,j} P_t/{d_{i,j}^\eta}}{\sigma^2 +
\sum_{k\in N, k\ne i}H_{k,j} P_t/d_{k,j}^\eta I_k} \ge \beta \right)
\end{equation}

Since the system activity is independent from slot to slot, by the
strong law of large numbers, we have, with probability one
\vspace{-1mm}
\begin{equation} \label{eqn:slln}
\lim_{t\to \infty} \frac{C_{i,j}(t)}{B_i(t)} = p_{i,j}
\end{equation}

The $p_{i,j}$ estimate thus obtained above is used by node $j$ as the
\emph{receive link weight} of the link $(i,j)$. We now analyse the
monotonicity of $p_{i,j}$ with $d_{i,j}$ for its applicability as
distance-like information.

\subsubsection{Monotonicity of $p_{i,j}$ with $d_{i,j}$ Under Isotropic Conditions}
From Equation~\ref{eqn:p_{i,j}}, we have
\vspace{-1mm}
\begin{eqnarray*}
p_{i,j}&=& (1-\alpha)\mathsf{Pr}\left\{\frac{H_{i,j} P_t/{d_{i,j}^\eta}}{\sigma^2 +
\sum_{k\in N} \left(H_{k,j} P_t/d_{k,j}^\eta\right) I_k - H_{i,j} P_t/{d_{i,j}}^\eta}\ge \beta \right\} \\
      &=& (1-\alpha)\mathsf{Pr}\left\{ (1+\beta) H_{i,j} P_t/d_{i,j}^\eta \ge \beta \left(\sigma^2 +
\sum_{k\in N} \left(H_{k,j} P_t/d_{k,j}^\eta\right) I_k \right) \right\} \\
&=& (1-\alpha)\mathsf{Pr}\left\{{\cal I}_j \le  \frac{(1+\beta) H_{i,j} P_t}{\beta d_{i,j}^\eta} - \sigma^2\right\}
\end{eqnarray*}
where ${\cal I}_j := \sum_{k\in N} \left(H_{k,j}
  P_t/d_{k,j}^\eta\right) I_k$ is the random variable that represents
the total power received at $j$.

For a large \emph{spatially homogeneous} network, where the node
density is constant over the entire region, the total power received
at each node ${\cal I}_j$ can be assumed to be identically distributed
at all node locations, with common CDF $F(.)$. This assumption will be
shown to be valid (to a good approximation) for the interior nodes
using simulations in \ref{sec:power_distribution}. Then, using $A(.)$,
the cumulative distribution function of $H_{i,j}$ (which is assumed to
be identical across all $(i,j)$), the equation above becomes
\vspace{-2mm}
\begin{equation} \label{eqn:monotonicity} p_{i,j}=
  (1-\alpha)\int_0^\infty F\left(\frac{(1+\beta) h P_t}{\beta
      d_{i,j}^\eta} - \sigma^2\right) \mbox{d}A(h)
\end{equation}
Since $F(.)$ and $A(.)$ are monotone increasing functions (as both are
CDFs), it can be seen from Equation~\ref{eqn:monotonicity} above that
$p_{i,j}$ is monotone decreasing with $d_{i,j}$. This allows us to
replace $d_{i,j}$ in the previous algorithm (in
Section~\ref{sec:algorithm using distances}) by $-p_{i,j}$, to obtain
a distance-free algorithm which is described next in
Section~\ref{sec:distance_free_algorithm_iso}.

\vspace{-1mm}
\subsection{DISCRIT: DIStributed CRITical geometric graph algorithm}
\label{sec:distance_free_algorithm_iso}
At the end of the Hello-protocol based neighbour discovery, each node $i$
has link weights $p_{j,i}$ for each of its neighbours $j$. Every node
$i$ maintains a \emph{$p$-threshold} $p(i)$ and an adjacent node list
$N(i)$. At any iteration, $N(i)$ is the set of nodes $j$ whose
$p_{j,i}$ values are greater than or equal to $p$-threshold $p(i)$.
\begin{enumerate}
\item \textbf{Initialisation:} For every node $i$, the $p$-threshold
$p^{(0)}(i)$ is initialised to the maximum link weight, and the
adjacent node list $N^{(k)}(i)$ contains only the node(s) with the
maximum weight. That is, for all $i \in N$,
\[p^{(0)}(i) = \max_j \{p_{j,i}\} \qquad \mbox{and} \qquad N^{(0)}(i)= \arg \max_j \{{p_{j,i}}\} \vspace{-1mm}\]
Set iteration index $k=0$.
\item \textbf{$p$-threshold unicast:} Every node $i$ informs its
current $p$-threshold $p^{(k)}(i)$ to all its current
\emph{adjacent nodes}, i.e., nodes in $N^{(k)}(i)$. Thus, the node $i$
also receives the $p$-thresholds $p^{(k)}(j)$ from some of its
neighbours given by the set $S^{(k)}(i) ~=~\{j:i \in N^{(k)}(j)\}$.
\item \textbf{Updating the Adjacent Node List:} The node then updates
its $p$-threshold $p^{(k+1)}(i)$ to the minimum of the $p$-thresholds
it received. The minimum finding includes the node's current
$p$-threshold $p^{(k)}(i)$ also. The adjacent node list $N^{(k+1)}(i)$
is also updated accordingly as the set of nodes whose $p_{j,i}$s are
greater than the updated $p$-threshold $p^{(k+1)}(i)$. Let
$t^{(k)}(i)= \min\{p^{(k)}(j): j\in S^{(k)}(i)\}$, which is the
smallest of the $p$-thresholds received by $i$. Then
\vspace{-1mm}
\[p^{(k+1)}(i) = \min\{p^{(k)}(i),t^{(k)}(i)\} \qquad \mbox{and} \qquad N^{(k+1)}(i) = \{j: p_{j,i} \geq p^{(k+1)}(i)\} \vspace{-1mm}\]
\item \textbf{Terminating Condition:}
The algorithm terminates if all the $p$-thresholds in an iteration
remain unchanged, i.e.,

IF $p^{(k+1)}(i) = p^{(k)}(i)$ for all $i \in N$, \hspace{5mm} Call $N(i)= N^{(k)}(i)$, \hspace{3mm}\textbf{STOP}

ELSE  \hspace{5mm} Set $k=k+1$, \hspace{3mm} go to   \textbf{Step 2}.

The distributed terminating condition is as described for the
algorithm based on distances, and can be used by the nodes to
terminate locally.
\item \textbf{Making links bidirectional:} Let $S(i) =\{j: i \in
  N(j)\}$. Thus $S(i)$ represents the nodes having $i$ as its adjacent
  node. The bidirectionality is achieved by updating the adjacent node
  list as $N(i) = N(i) \cup S(i)$. The graph resulting from the
  algorithm $\hat{{\cal G}}_1$ is then given by $\hat{{\cal G}}_1 =
  (\mathbf{V},\hat{E}_1)$ where $\hat{E}_1=\{(i,j): i\in N, j\in
  N(i)\}\hfill \qed$
\end{enumerate}

\begin{remarks}
Unlike distances $d_{i,j}$, $p_{i,j} \ne p_{j,i}$ in
general. Therefore, the graph obtained after the terminating condition
in Step~\textbf{4} need not be bidirectional. Hence, additional edges
are added to the graph in Step~\textbf{5} to ensure
bidirectionality.
\end{remarks}

\begin{remarks}
  Note that for DISCRIT to be valid, we need \emph{spatial
    homogeneity} for $p_{i,j}$ monotonicity as well as the Penrose's
  result (Corollary~\ref{cor:relation between r_1 and r_crit}) to
  hold. While randomised lattice and grid deployments are spatially
  homogeneous, a uniform i.i.d. deployment, in general, can create
  \emph{sparse} and \emph{dense} node placements. However, as $n$ is
  increased, uniform i.\ i.\ d.\ deployment is homogeneous w.h.p.\ in
  a sense described below. Further, Corollary~\ref{cor:relation
    between r_1 and r_crit} is known to be true for uniform i.\ i.\
  d.\ deployment. We will use simulations to study the applicability
  to other deployments.

  Given a uniform i.\ i.\ d.\ deployment $\mathbf{V}$, any $r>0$, and
  any point $x$ within the region ${\cal A}$, define \emph{closed
    disc} of radius $r$ around $x$, $D_r(x):=\{y \in
  \mathbf{R}^2: \parallel y-x \parallel \le r\}$. Let
  $N_r(x;\mathbf{V})$ be the number of nodes in $D_r(x)$, i.e., within
  a radius of $r$ around $x$. Define the \emph{interior} of ${\cal A}$
  as $\tilde{\cal A}(r) = \{x \in {\cal A}: D_r(x) \subset {\cal
    A}\}$.  The following theorem gives the joint convergence of
  $N_r(x;\mathbf{V})$ for all $x \in \tilde{\cal A}(r)$.
\begin{theorem} \label{thm:homogeneity}
For any $\epsilon >0$ however small,
\[\lim_{n \to \infty} {\cal P}^n \left\{ \mathbf{V}: \frac{n}{\mid {\cal A} \mid} (1-\epsilon) \le \frac{N_r(x;\mathbf{V})}{\pi r^2} \le \frac{n}{\mid {\cal A} \mid} (1+\epsilon) \mbox{ for every } x \in \tilde{\cal A}(r) \right\} = 1\]
\end{theorem}
\begin{IEEEproof}
See Appendix~\ref{appendix:proof of homogeneity}
\end{IEEEproof}
The result above implies that, for a uniform i.\ i.\ d.\ placement of
nodes, the node density around every \emph{interior} point, i.e.,
$\frac{N_r(x;\mathbf{V})}{\pi r^2}$, is arbitrarily close to the
network density $\frac{n}{\mid {\cal A} \mid}$ w.h.p., thus implying
the homogeneity of dense uniform i.\ i.\ d.\ deployments.
\end{remarks}

\begin{remarks} \label{rmk:edge_effect} For finite regions, even when
  the deployment is spatially homogeneous, there will be an
  \textbf{edge effect}. The received power at a periphery node is
  usually less compared to the receive power at the nodes in the
  interior, because of the difference in node density at the periphery
  and the centre of the region (non-homogeneity at the
  edges)~\footnote{Note that even the homogeneity result in
    Theorem~\ref{thm:homogeneity} is applicable at the \emph{interior}
    of the region, and not at the edges.}. This \emph{edge effect}
  will distort the behaviour of DISCRIT, as will be seen in the
  simulation results in Section~\ref{sec:simulation}. A remedy for
  \emph{edge effect} is to extend the node deployment beyond the
  boundaries of the area of interest, so that the actual region of
  interest does not experience the edge effect.
\end{remarks}

\section{DISCRIT:Simulation Results} 
\label{sec:simulation}

\subsection{Power Distribution}
\label{sec:power_distribution}
In Section~\ref{sec:distance_free_algorithm_iso}, we assumed that the
total power at a receiver node (Signal~+~Interference~+~Noise) is
identically distributed at every receiver. When the number of nodes is
sufficiently large, a receiver far away from the edges sees almost the
same concentration of nodes around it and hence experiences the same
distribution of total received power. This is particularly so if the
exponent for power loss with distance $\eta$ is large, because only
the nearby nodes can make a measurable difference to the power
received. This also reduces the ``edge effect'' seen by the receivers
close to the edges.

\begin{figure}[ht]
\begin{center}$
\begin{array}{ccc}
  \includegraphics[width= 5.8cm, height= 5.3cm]
    {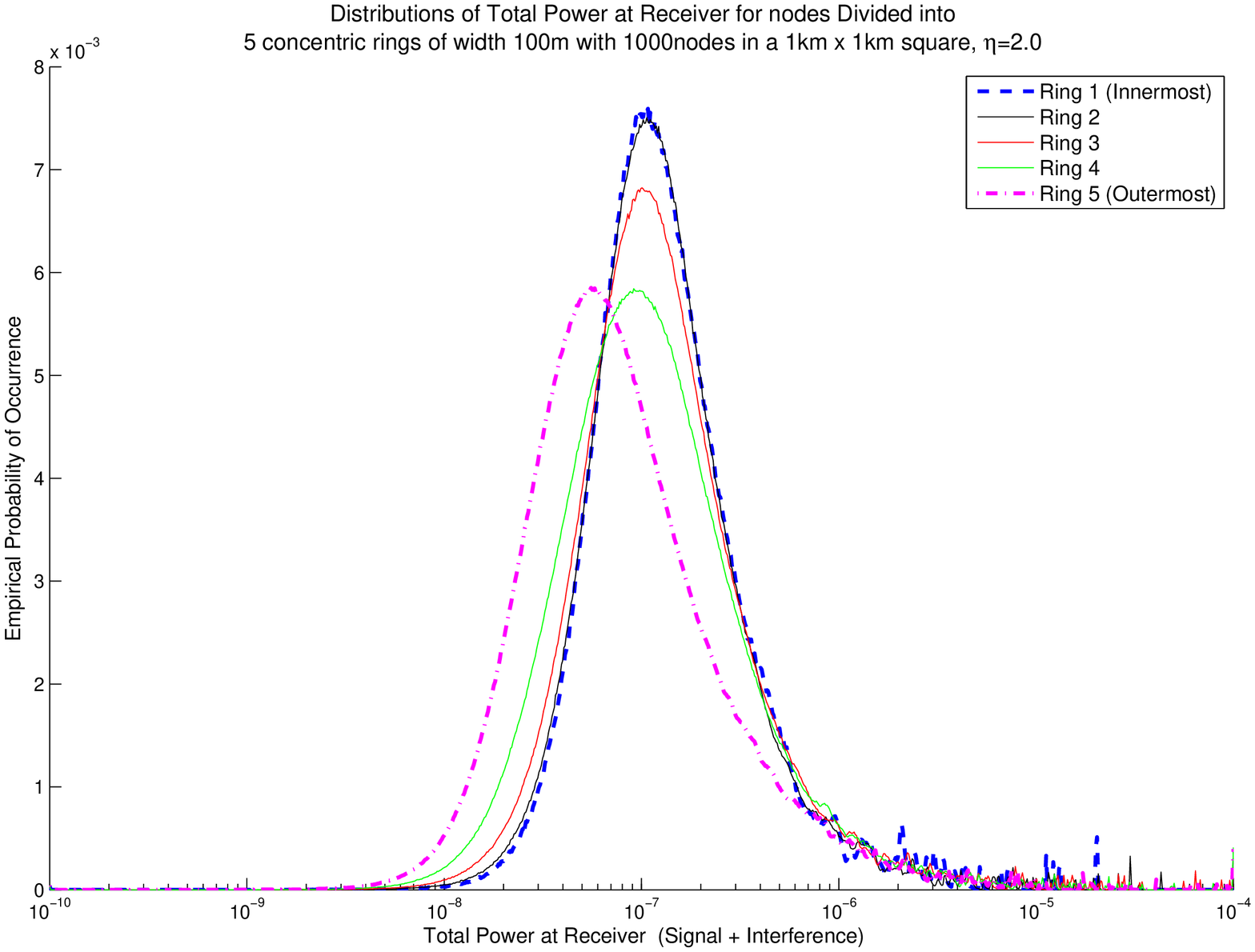}&
  \includegraphics[width= 5.8cm, height= 5.3cm]
    {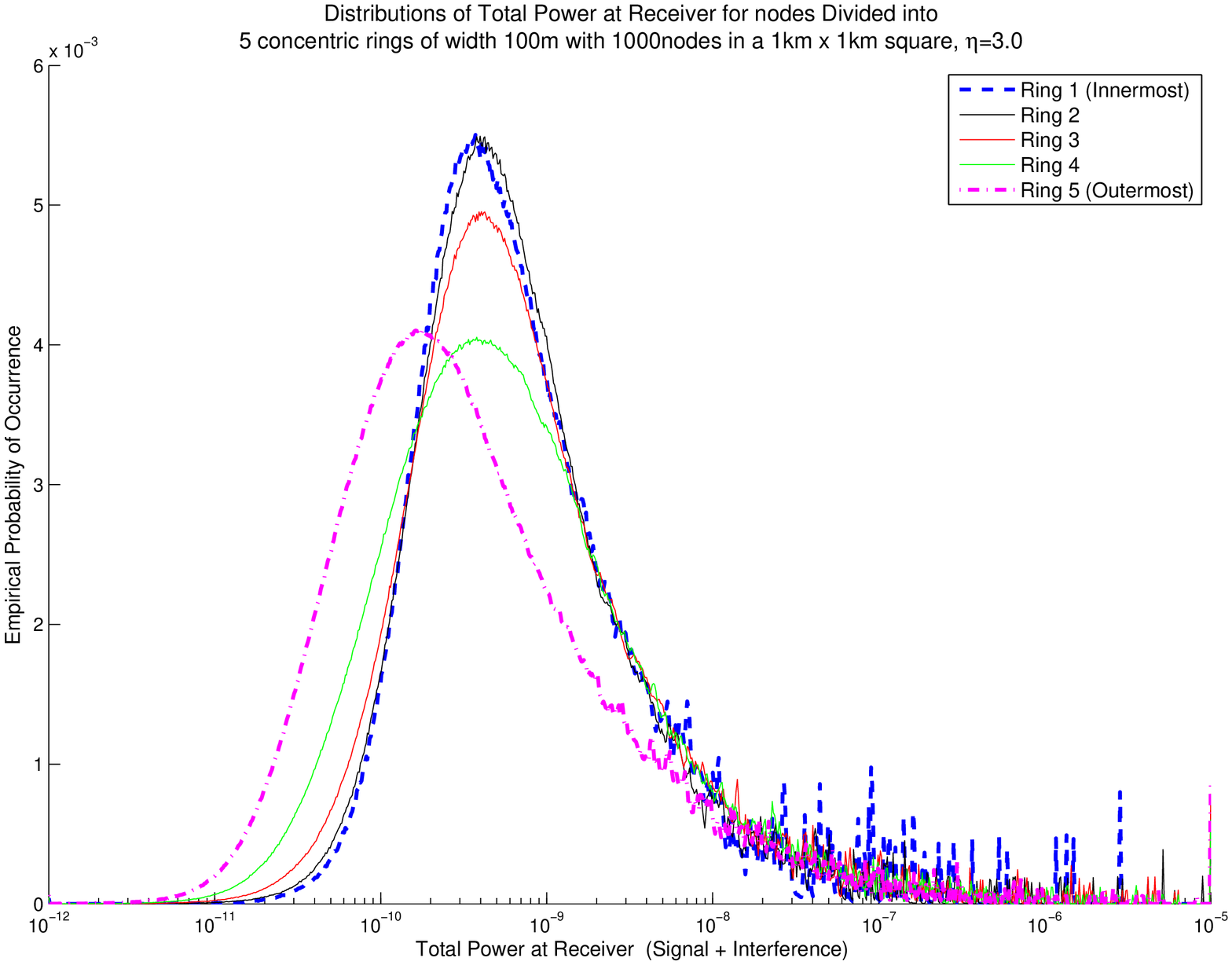}&
  \includegraphics[width= 5.8cm, height= 5.3cm]
    {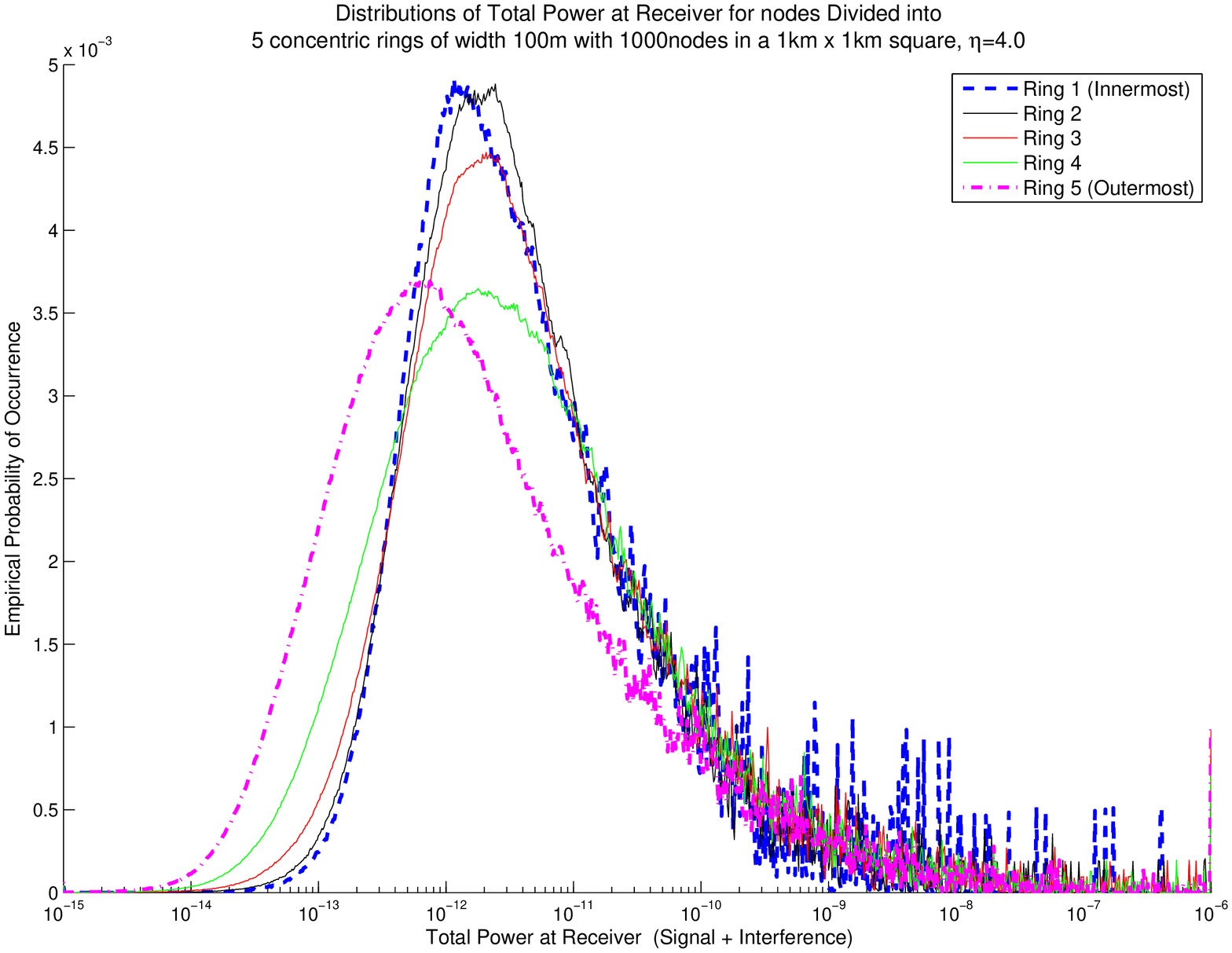}\\

  \includegraphics[width= 5.8cm, height= 5.3cm]
    {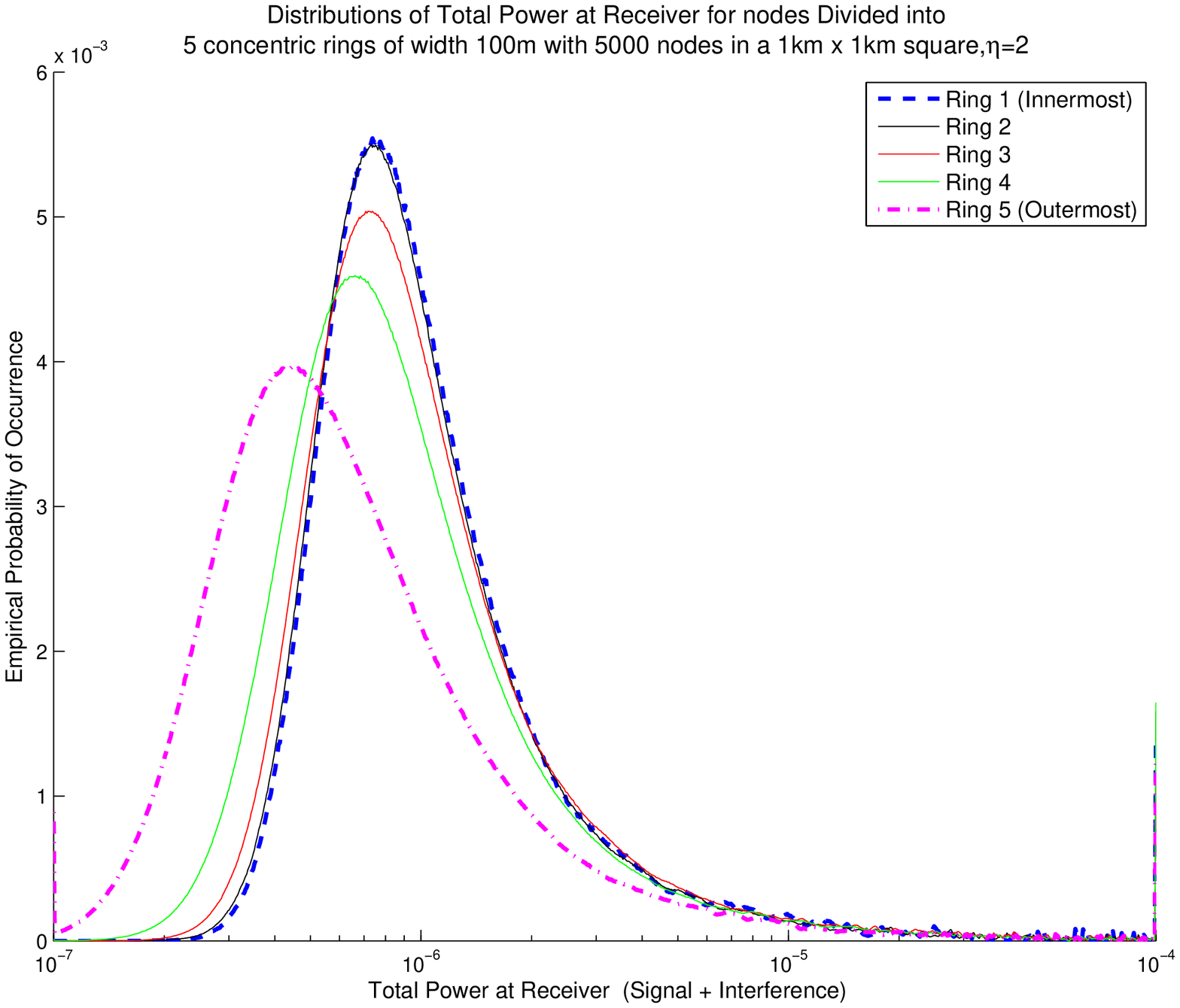}&
  \includegraphics[width= 5.8cm, height= 5.3cm]
    {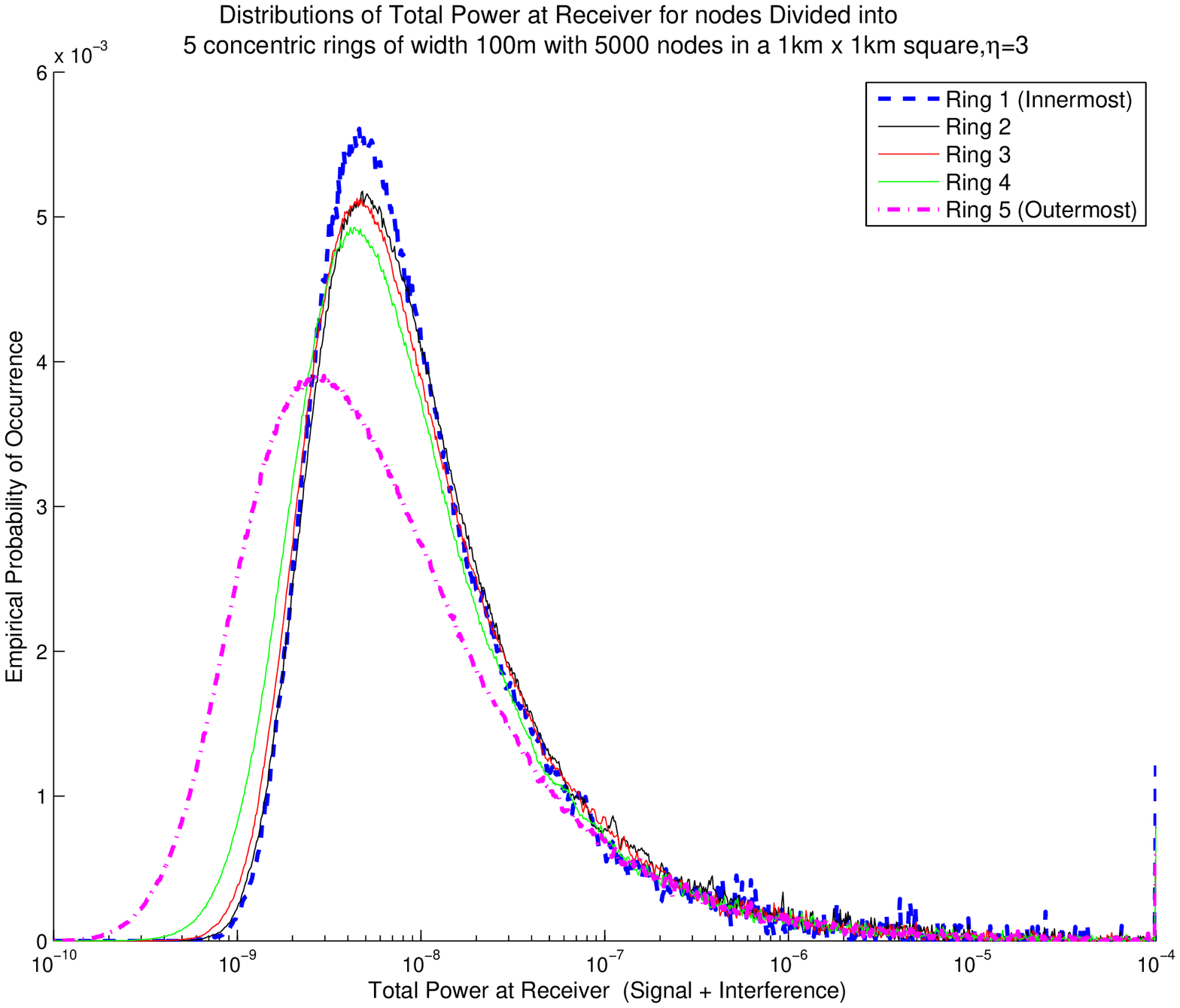}&
  \includegraphics[width= 5.8cm, height= 5.3cm]
    {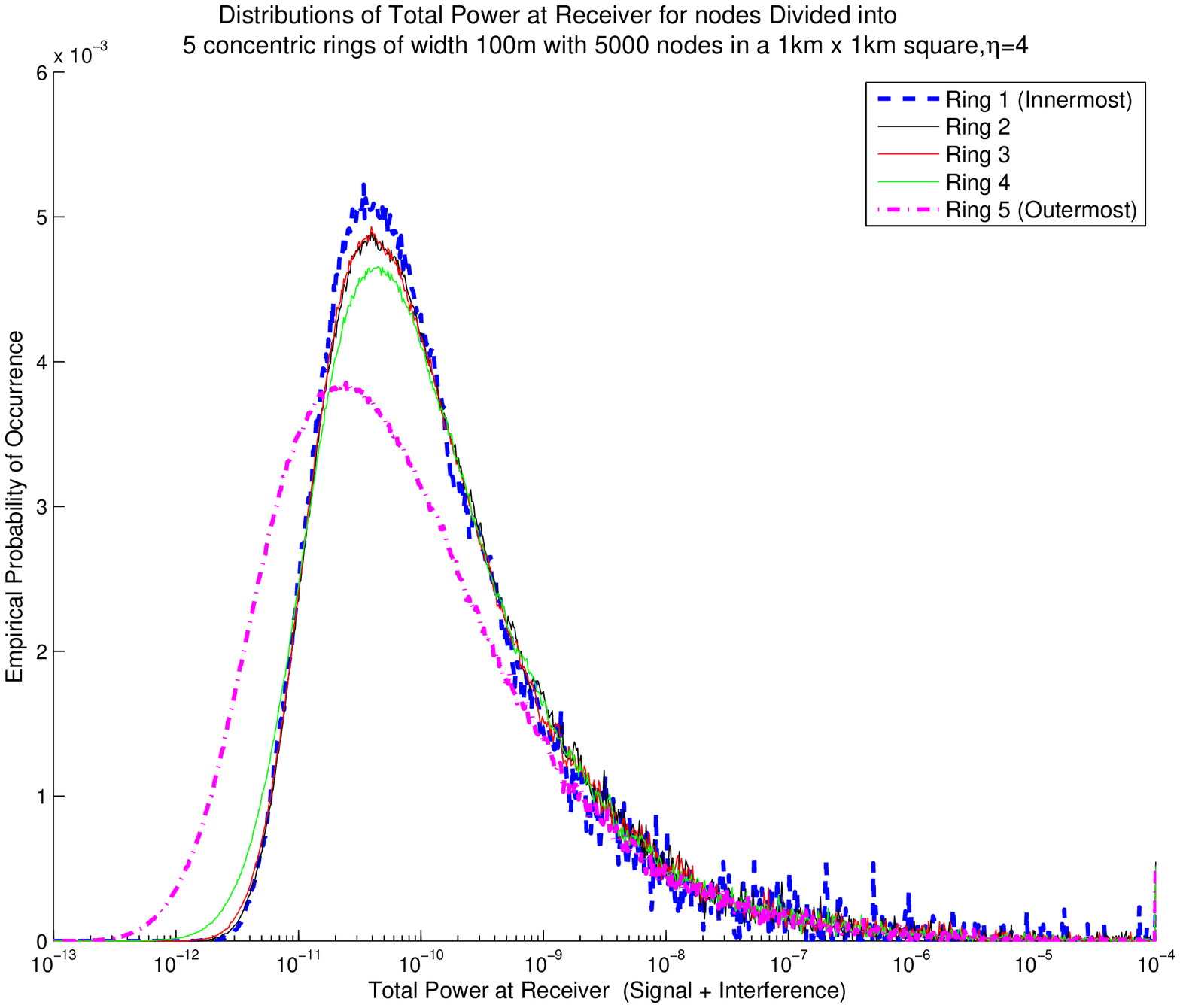}\\
\end{array}$
\end{center}
\caption{Empirical histograms of total received power at nodes in 5
  concentric annular regions in a 1~Km $\times$ 1~Km region, with the
  nodes distributed independently and uniformly over the region.  The
  top row corresponds to 1000 nodes, and the bottom row to 5000 nodes.
  The columns in this $2 \times 3$ array of plots correspond
  successively to path loss exponents $\eta = 2, 3,4$.}
\label{fig:power_distribution}
\end{figure}

For a verification of this in simulation, two random deployments of
nodes in a unit square were considered, one with 1000 nodes and the
other with 5000 nodes. $\eta$ is also varied from 2.0 to 4.0. In each
case, a large number of time slots of the slotted-Aloha protocol were
simulated in Matlab, where each node transmits with some probability
$\alpha$ independent of all other nodes, and acts as a receiver
otherwise. The nodes in the unit square are also divided into 5
concentric annular regions of equal width (of 0.1~units).  The total
power received at the nodes in each region was calculated and an
empirical probability of occurence of each value of power (upto a
certain resolution) was calculated. The resulting empirical
distribution of powers is shown in
Figure~\ref{fig:power_distribution}.

We notice from Figure~\ref{fig:power_distribution} that, for a given
number of nodes (1000 or 5000), the assumption of the same received
power distribution across nodes becomes better as the path loss
exponent, $\eta$, increases, and as the nodes are taken farther away
from the edge of the region. The assumption also works better for a
larger node density. In fact, we see that for 5000 nodes, distributed
independently and uniformly over the region, and $\eta = 4$, the
approximation is excellent for nodes lying at points greater than
0.1~units from the boundary.


\subsection{Performance of DISCRIT}
\label{sec:simulation performance DISCRIT}
Simulations were carried out in Matlab using both random and uniform
(randomised lattice) deployments of 1000 nodes in a unit square
region. In both cases, the DISCRIT algorithm was performed on the
deployment under isotropic conditions and the results were compared
with the respective Critical Geometric Graphs for the deployments.

\begin{figure}[t]
\begin{center}$
\begin{array}{ccc}
  \includegraphics[width= 5.8cm, height= 5.3cm]
    {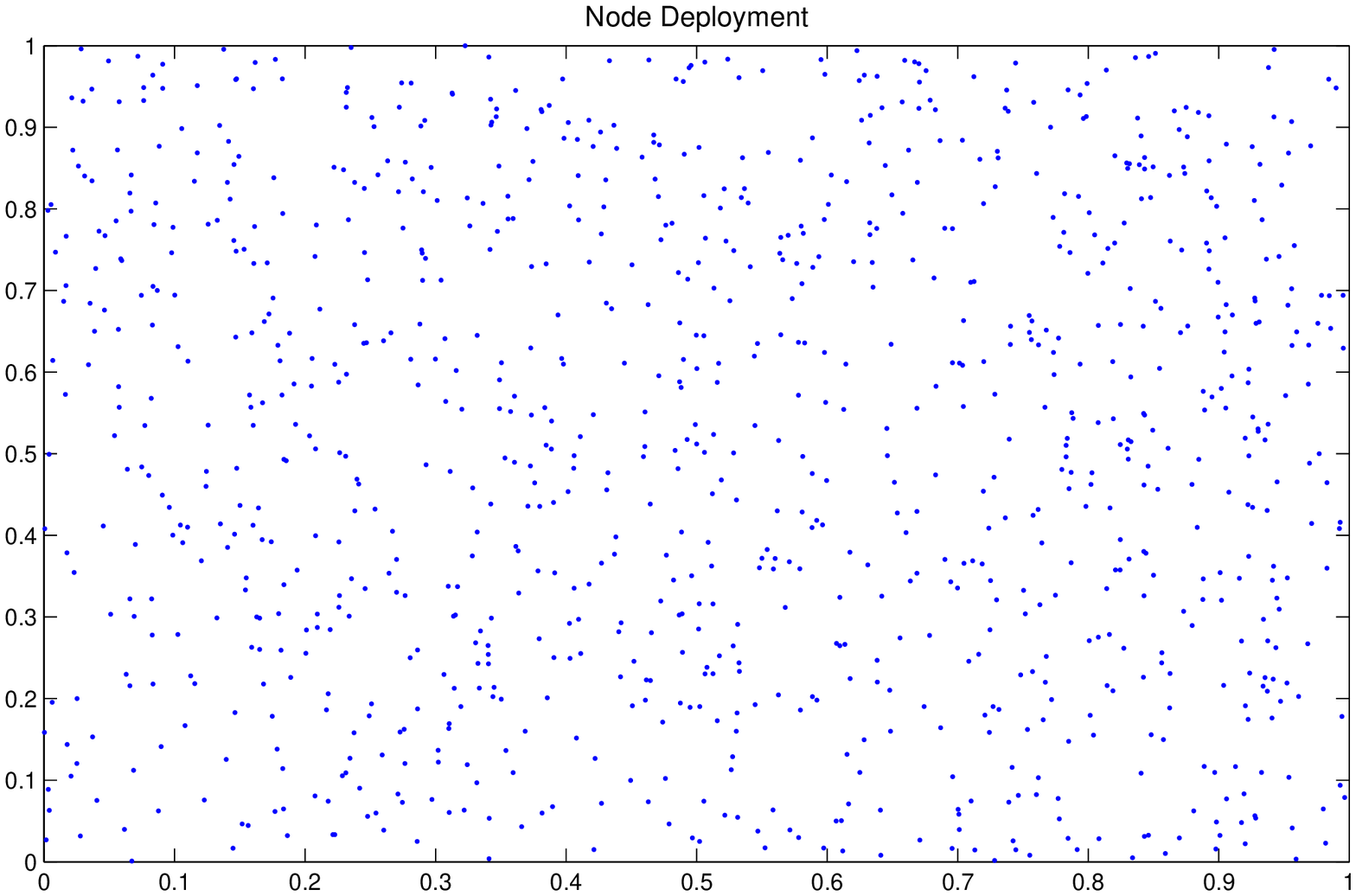}&
  \includegraphics[width= 5.8cm, height= 5.3cm]
    {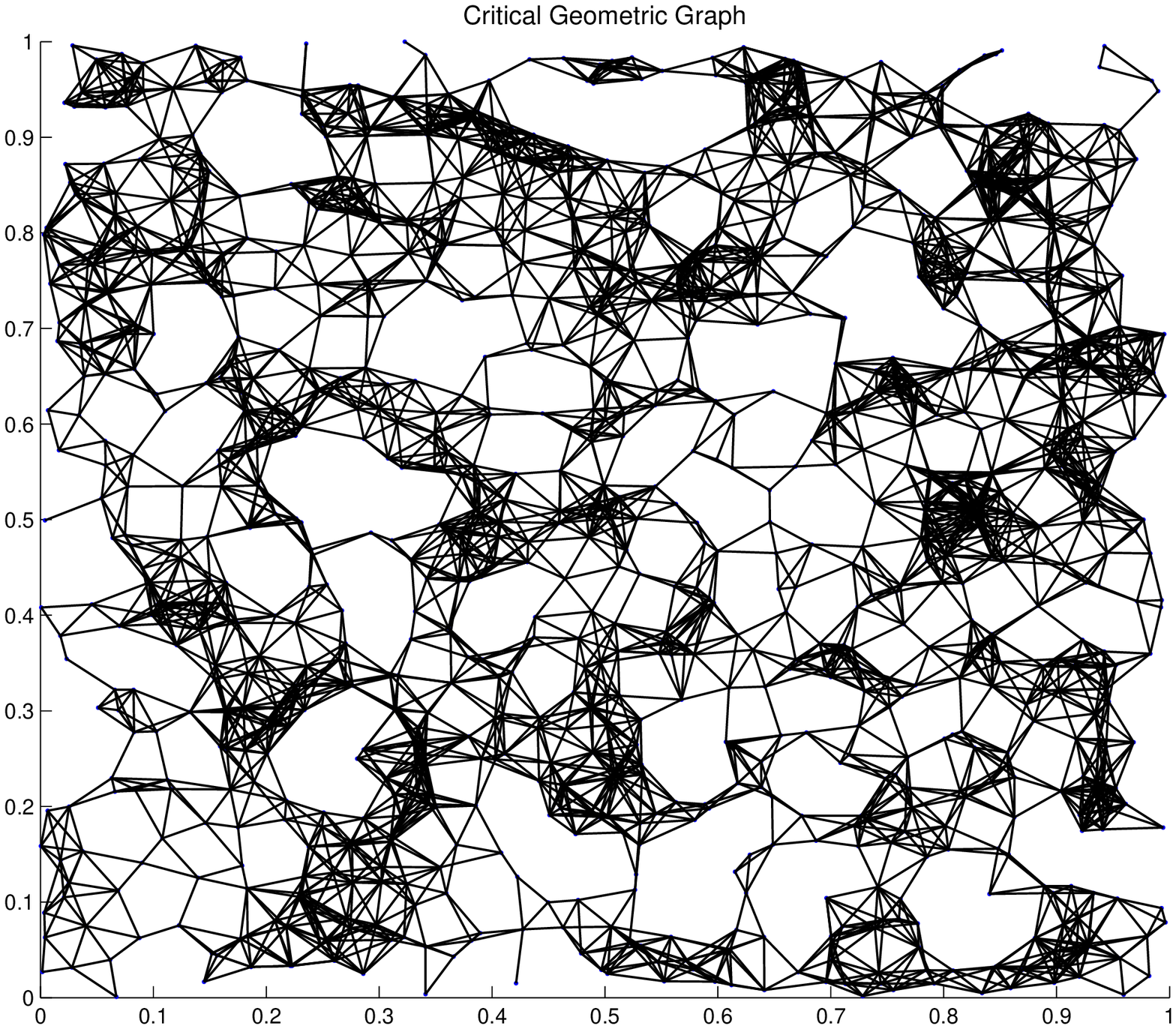}&
  \includegraphics[width= 5.8cm, height= 5.3cm]
    {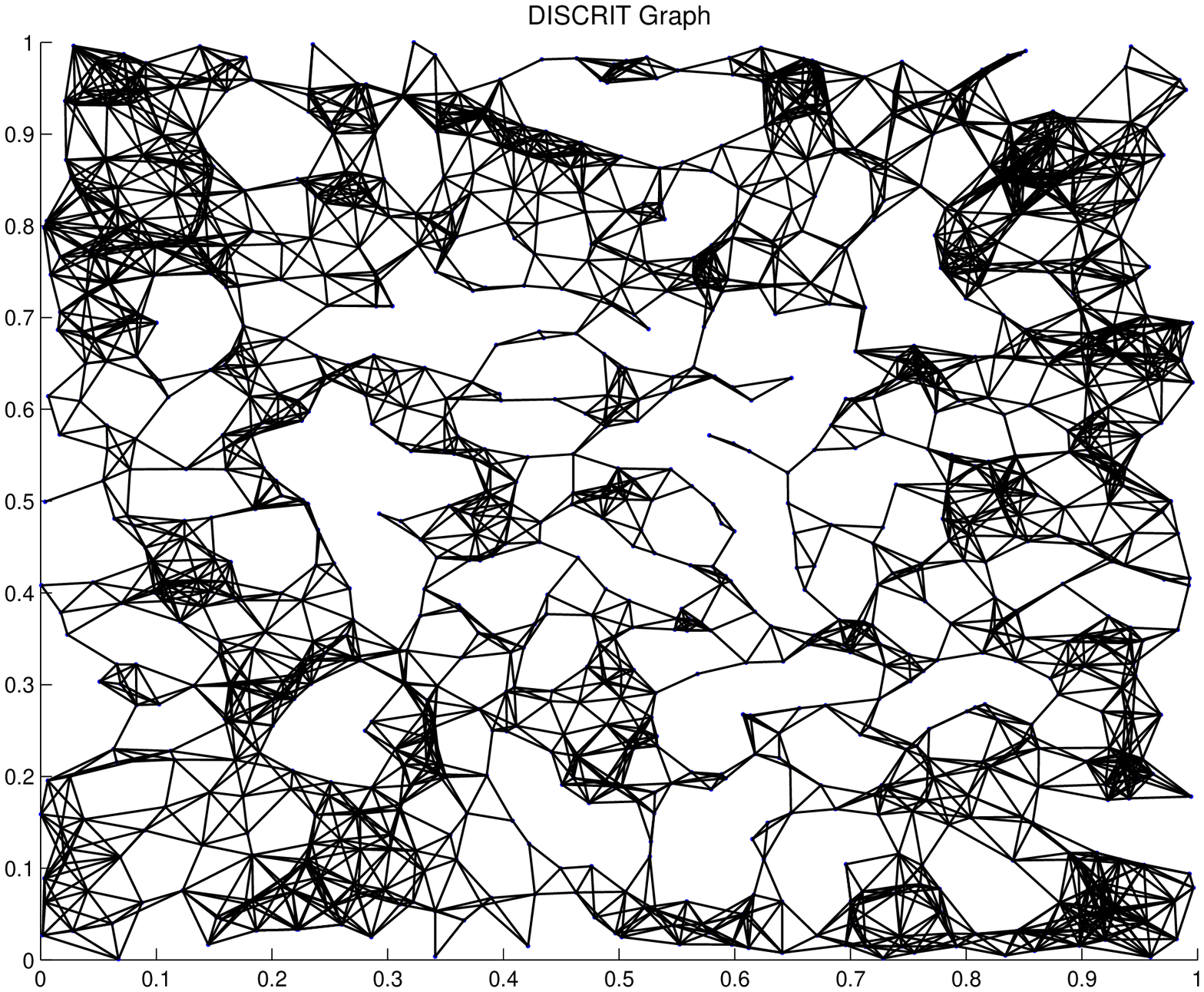}\\
  \includegraphics[width= 5.8cm, height= 5.3cm]
    {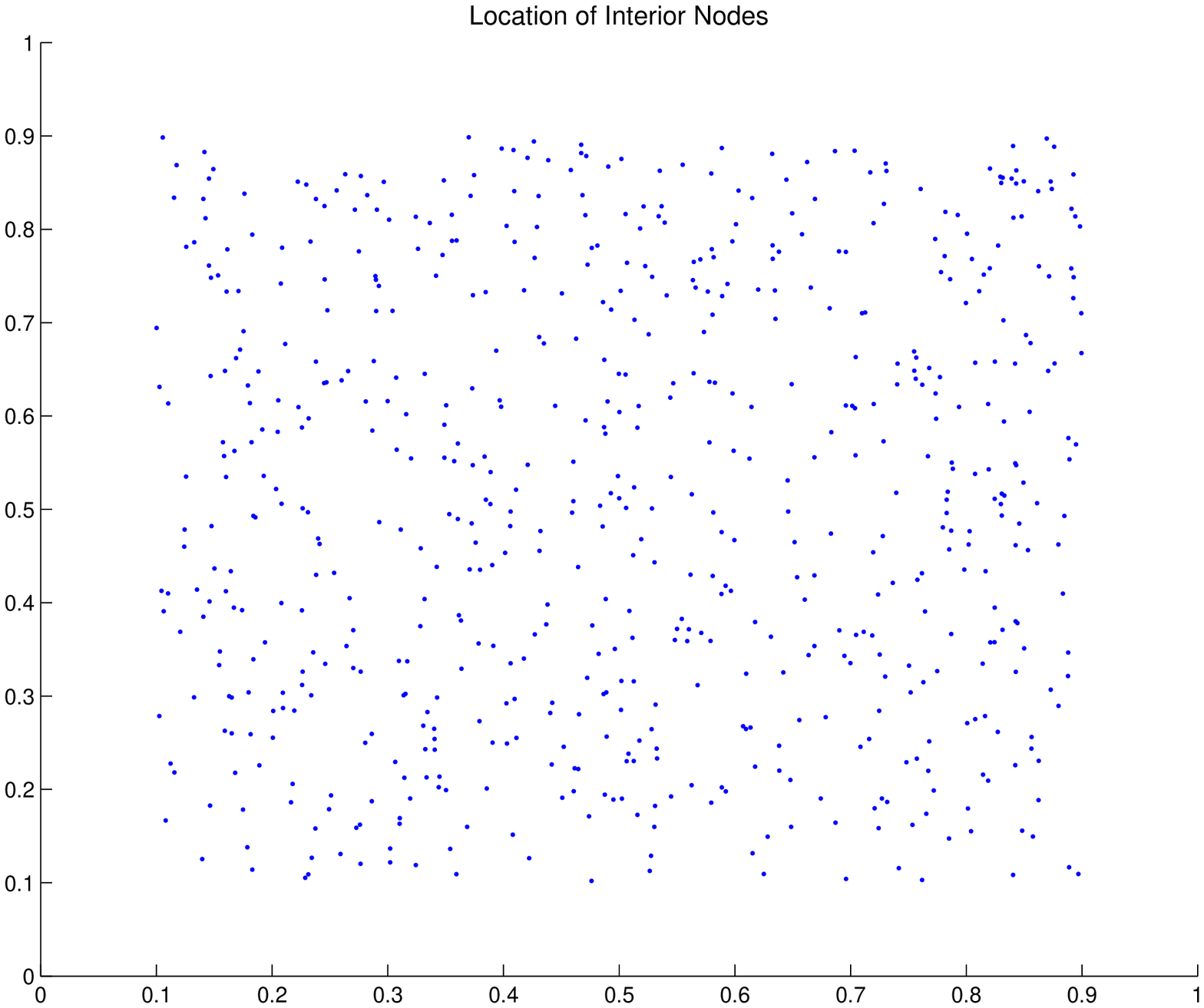} &
  \includegraphics[width= 5.8cm, height= 5.3cm]
    {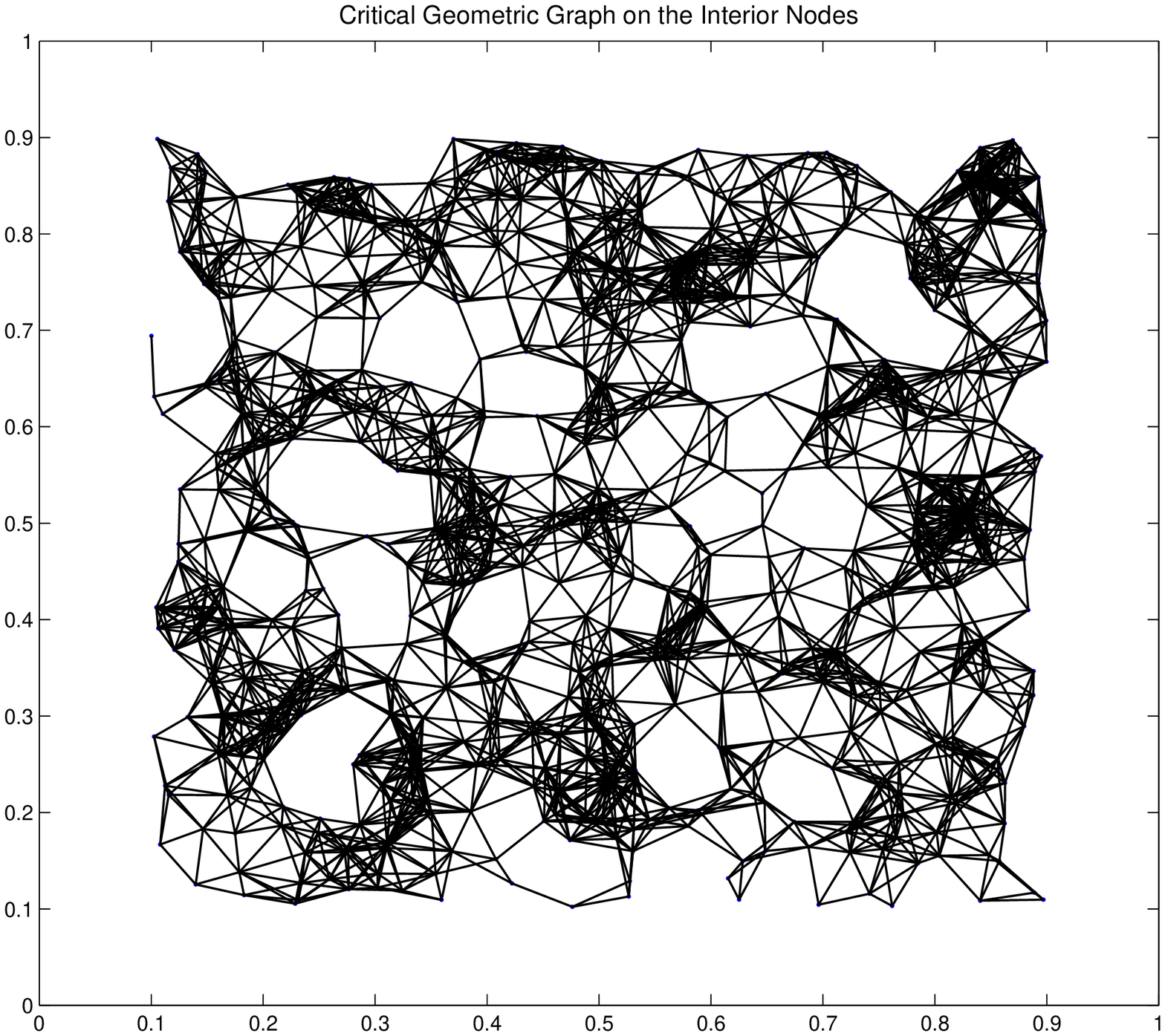}&
  \includegraphics[width= 5.8cm, height= 5.3cm]
    {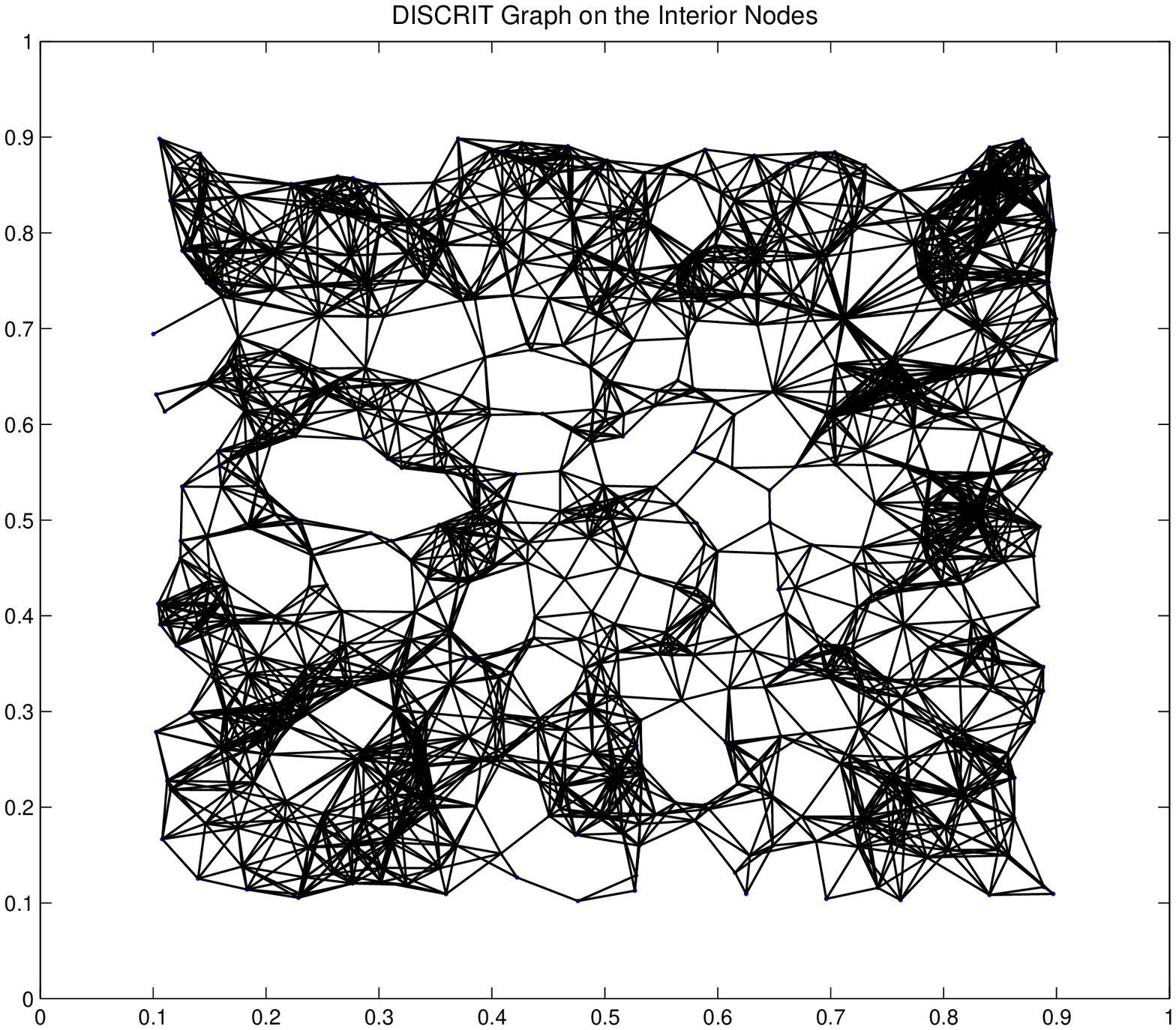}
\end{array}$
\end{center}
\caption{Uniform i.i.d.\ deployment: Comparison of the critical
  geometric graph and the approximation provided by DISCRIT. The
  leftmost column of plots shows the node locations, the middle column
  the critical geometric graph, and the rightmost column the result
  obtained from DISCRIT. In the bottom row of plots the algorithm is
  run only on the ``interior'' nodes shown in the leftmost plot. }
\label{fig:random_comparison_graphs}
\end{figure}
\begin{figure}[t]
\begin{center}$
\begin{array}{ccc}
  \includegraphics[width= 5.8cm, height= 5.3cm    ]
    {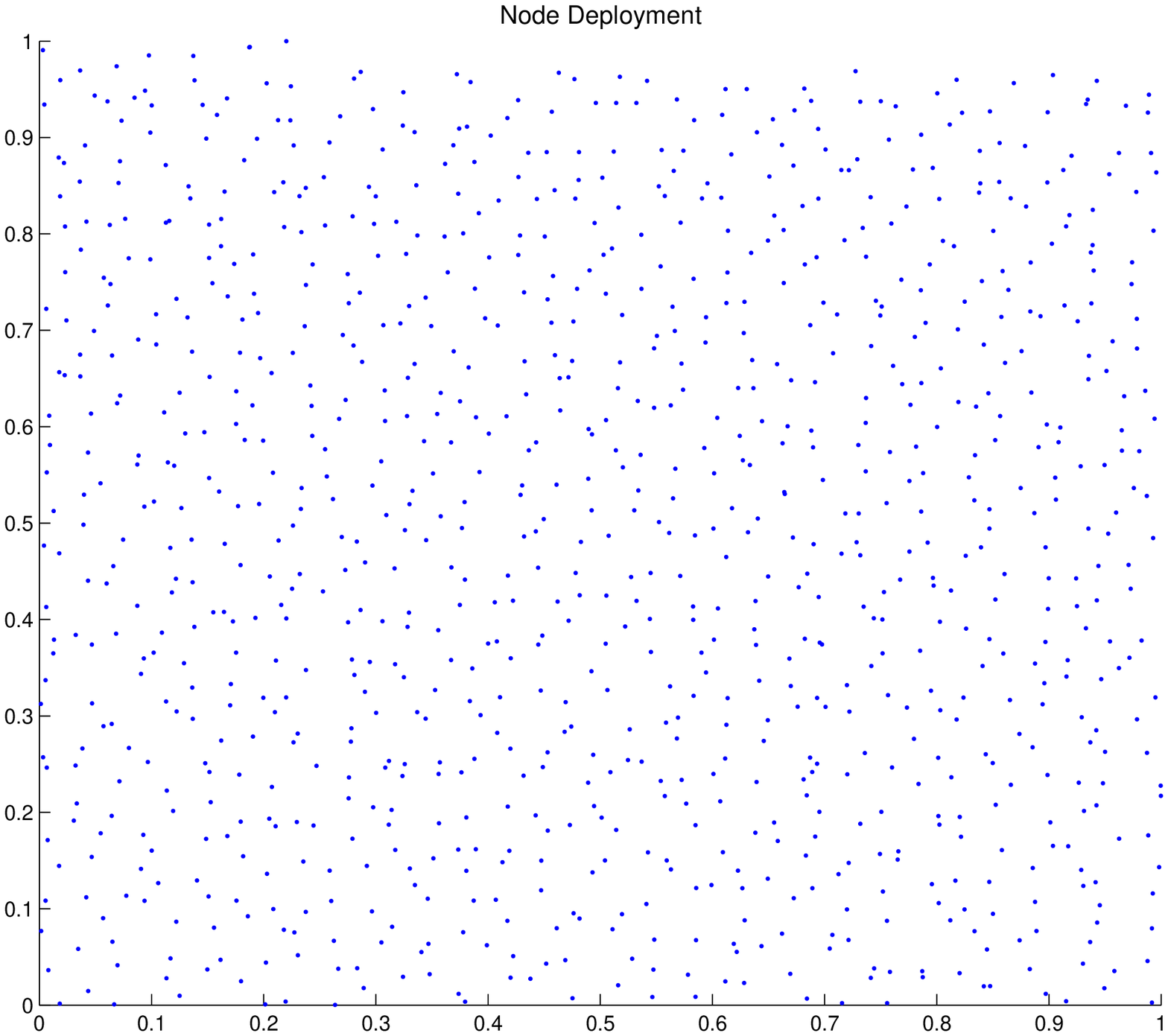}&
  \includegraphics[width= 5.8cm, height= 5.3cm    ]
    {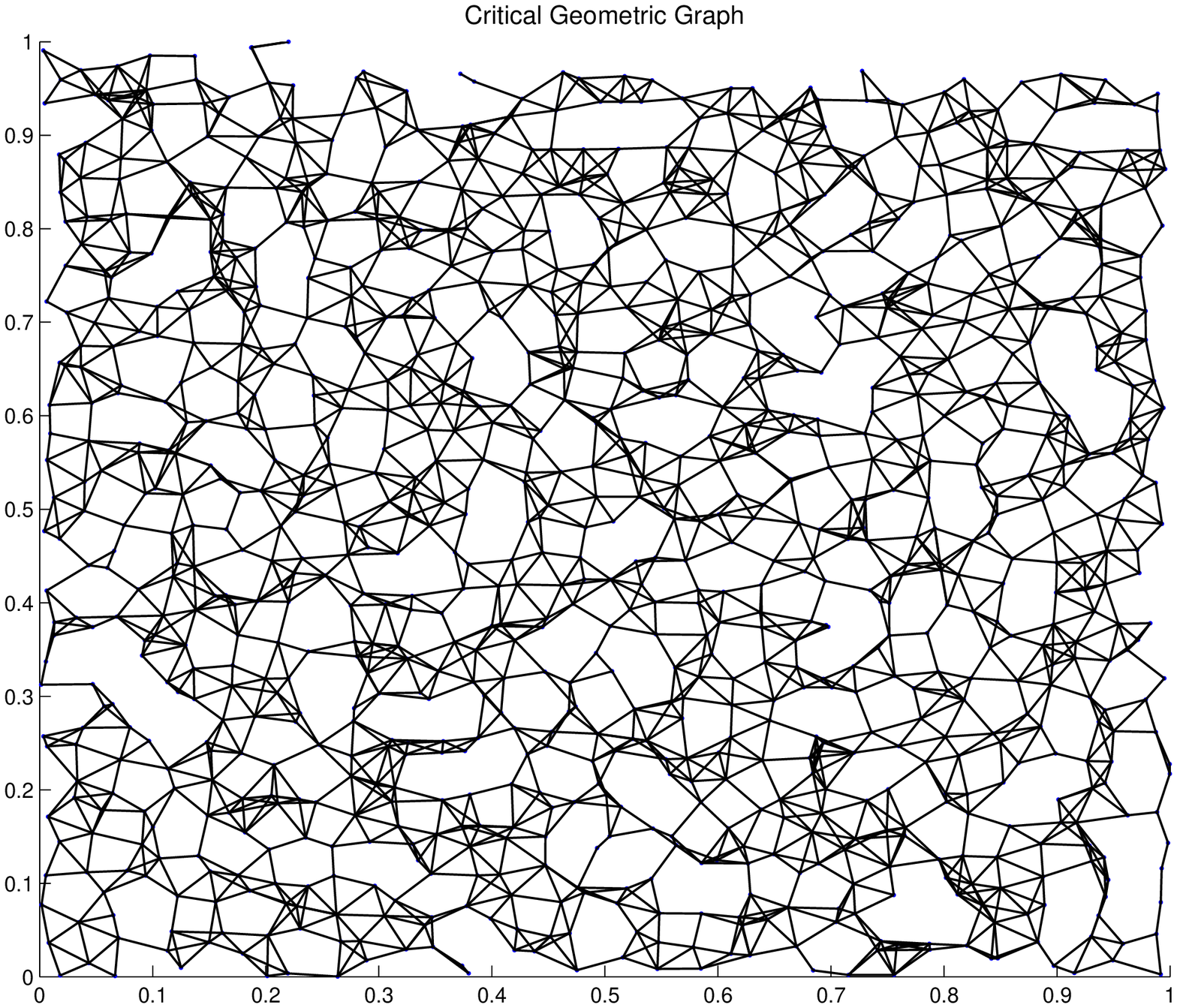}&
  \includegraphics[width= 5.8cm, height= 5.3cm    ]
    {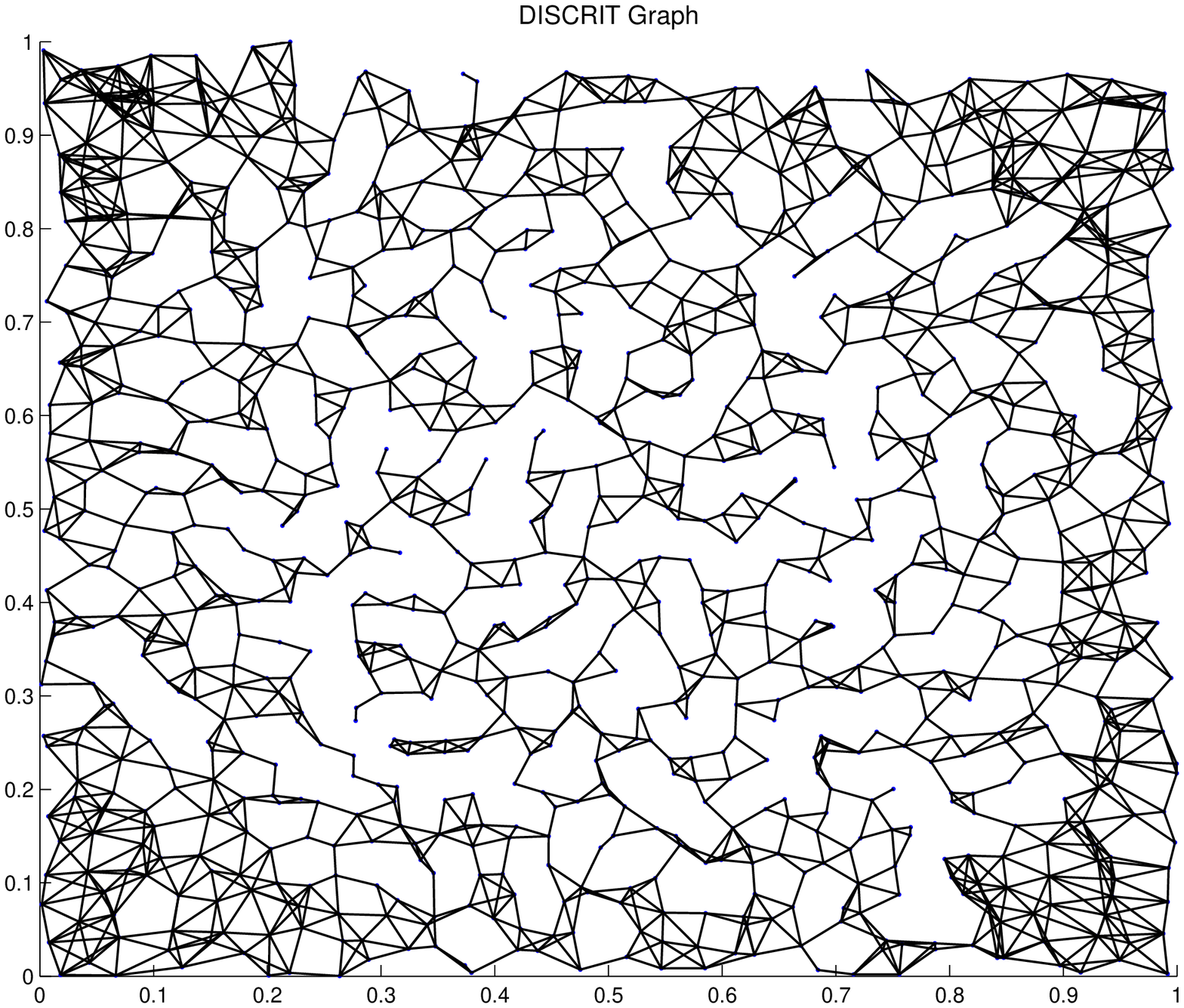}\\

  \includegraphics[width= 5.8cm, height= 5.3cm    ]
    {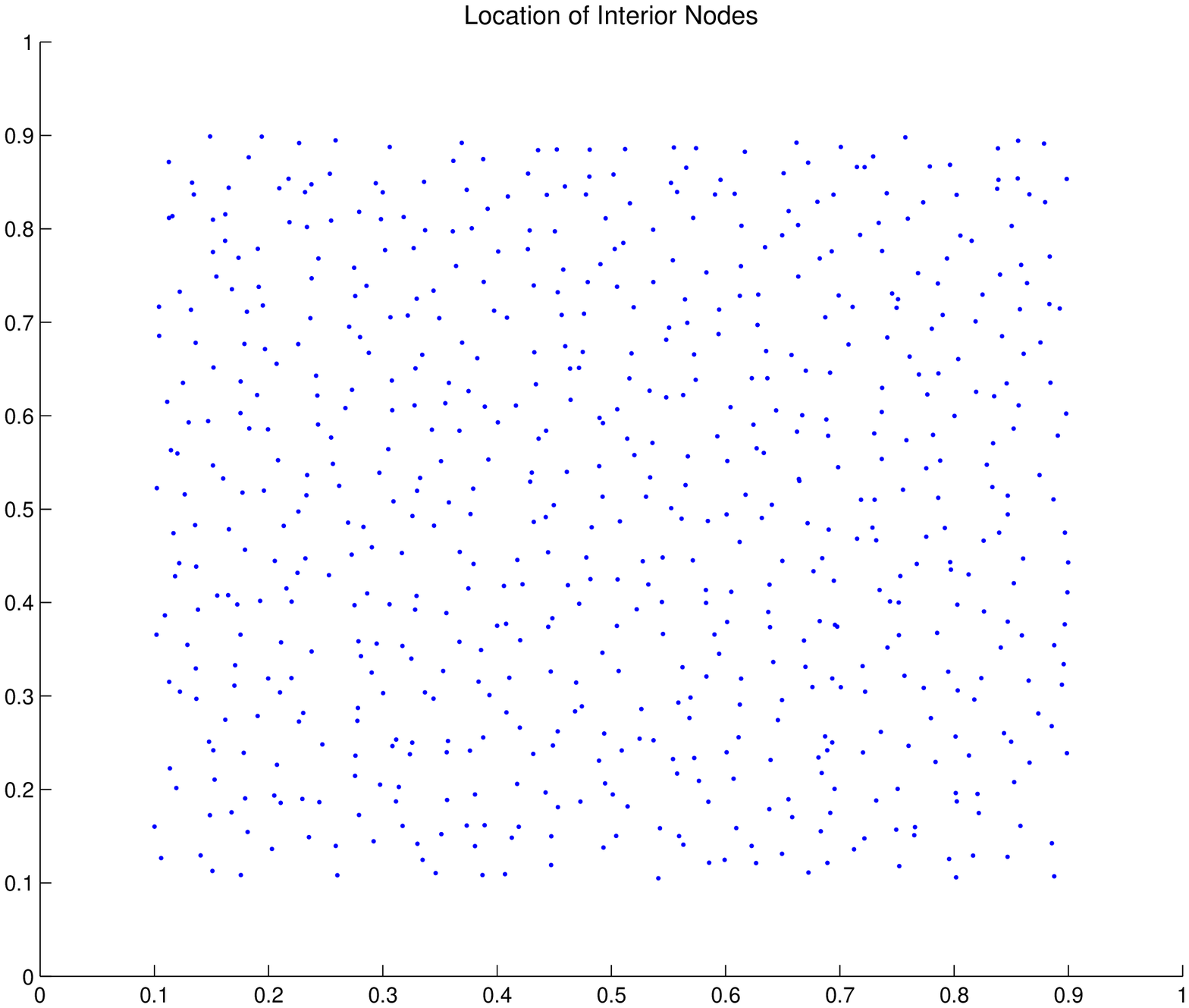} &
  \includegraphics[width= 5.8cm, height= 5.3cm    ]
    {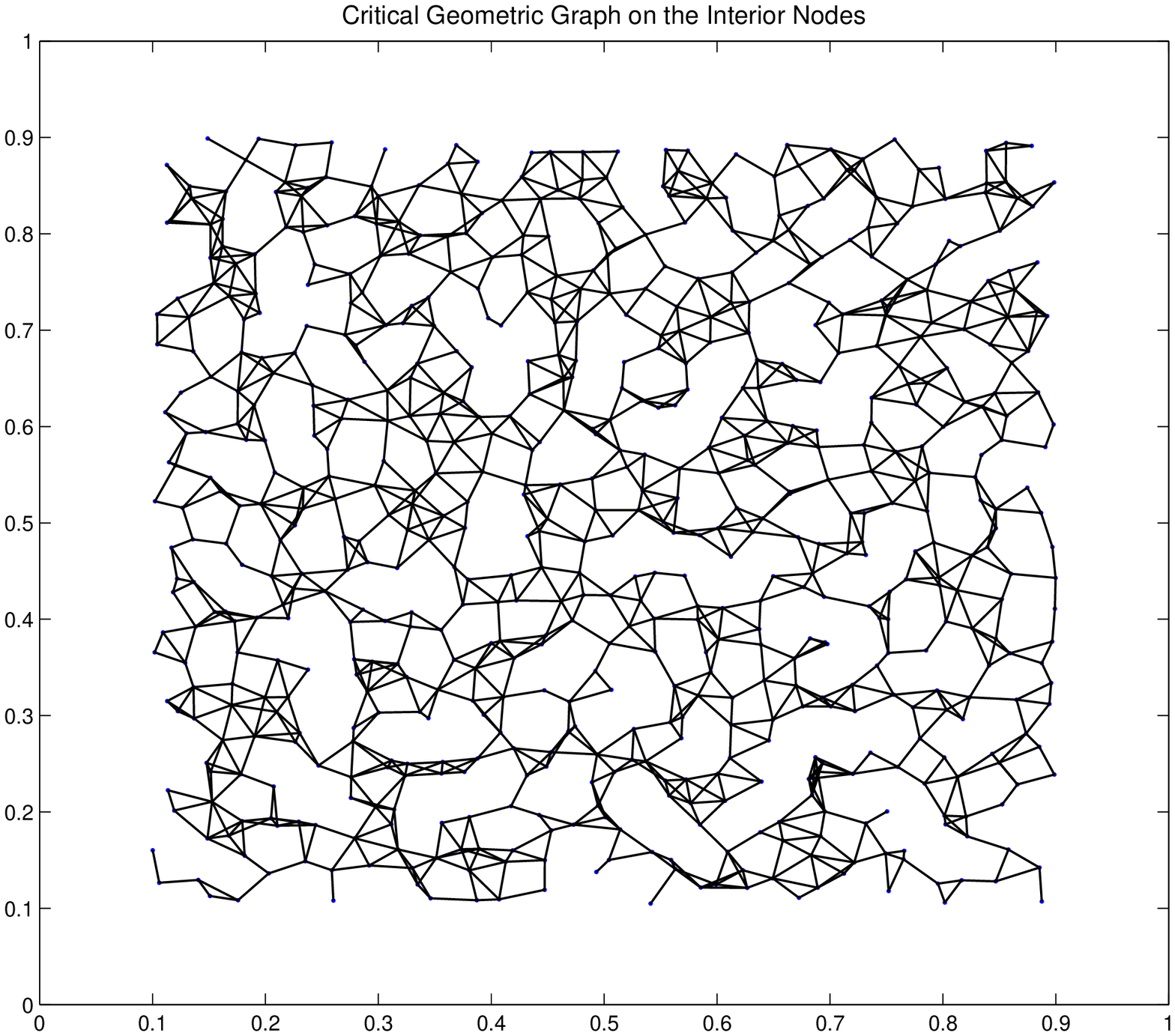}&
  \includegraphics[width= 5.8cm, height= 5.3cm    ]
    {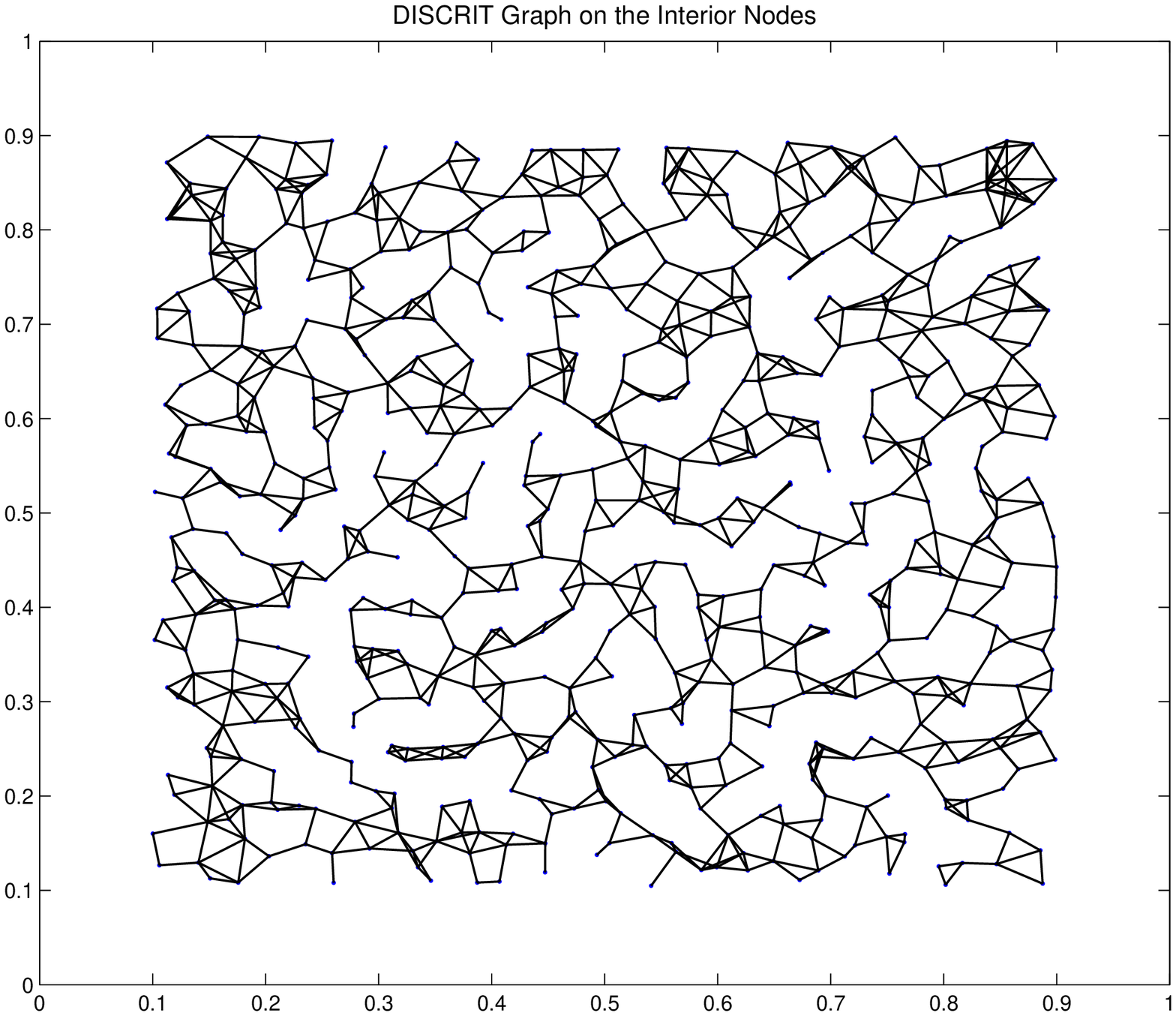}
\end{array}$
\end{center}
\caption{Randomised lattice deployment: Comparison of the critical
  geometric graph and the approximation provided by DISCRIT. The
  leftmost column of plots shows the node locations, the middle column
  the critical geometric graph, and the rightmost column the result
  obtained from DISCRIT. In the bottom row of plots the algorithm is
  run only on the ``interior'' nodes shown in the leftmost plot. }
\label{fig:uniform_comparison_graphs}
\end{figure}

Figure~\ref{fig:random_comparison_graphs} provides results for uniform
i.i.d.\ deployment.  From a visual comparison of the actual CGG and
the graph provided by DISCRIT we conclude that DISCRIT provide a graph
with a similar visual structure, though with fewer links (we will
evaluate this quantitatively below).  In the light of the discussion
about ``edge effects'' in Remarks~\ref{rmk:edge_effect}, the nodes
that were less than 0.1~units away from any edge of the unit square
were removed and all the preceding analysis was performed on the
interior node deployments, using the same link weights from Hello
protocol for the entire deployment. The resulting DISCRIT output graph
and CGG can be seen to be visually much closer to the actual CGG on
the interior nodes. Similarly,
Figure~\ref{fig:uniform_comparison_graphs} provides the results from
DISCRIT for the randomised lattice deployment.

\subsubsection{A quantitative measure of similarity between the CGG
  and output of DISCRIT}
We introduce a two-part measure to estimate the similarity between the
approximate GG given by DISCRIT, $\hat{\mathcal G}_1$ and the true
critical geometric graph, $\mathcal G_{crit}$ or the FNNGG
$\mathcal{G}_1$.  A geometric graph (GG) of radius $r$ is defined by
two criteria :

\begin{enumerate}
\item No edges of length more than $r$ should be present.
\item All edges of length at most $r$ should be present.
\end{enumerate}

Accordingly, we have defined the function $D(.)$ on a pair of graphs
$G_1$ and $G_2$ as $D(G_1,G_2)=\frac{|E_1\bigcap E_2^c|}{|E_1|}$,
where $E_i$ is the edge set of graph $G_i$, for $i=1,2$.
$D(\hat{\mathcal G}_1, \mathcal G_{crit})$ is the fraction of edges of
$\hat{\mathcal G}_1$ that are longer than $r_{crit}$ and $D(\mathcal
G_{crit}, \hat{\mathcal G}_1)$ is the fraction of edges shorter than
$r_{crit}$ that are missing in $\hat{\mathcal G}_1$. Ideally, both
these quantities should be $0$.

\begin{table}[t]
\centering
\begin{tabular}[t]{||c|c|c|c|c|c|c||}	\hline \hline
\multirow{2}{*}{\emph{Deployment}} &\multirow{2}{*}{\emph{$G_1$}}  &\multirow{2}{*}{\emph{$G_2$}}	&\multicolumn{2}{|c|}{\emph{$D(G_1,G_2)$}} &\multicolumn{2}{|c||}{\emph{$D(G_2,G_1)$}}
 \\	\cline{4-7}
& & &\emph{All Nodes} &\emph{Interior Nodes} &\emph{All Nodes} &\emph{Interior Nodes} \\ \hline\hline

\multirow{2}{*}{i.i.d} &$\hat{\mathcal{G}}_1$ & $\mathcal{G}_{crit}$ & 0.0610 &0.1248 &0.1656 &0.0793\\ \cline{2-7}
&$\hat{\mathcal{G}}_1$ &$\mathcal{G}_1$ &0.0610 & 0.1248 & 0.1656 & 0.0793	\\ \hline
\multirow{2}{*}{Randomised Lattice}  &$\hat{\mathcal{G}}_1$ &$\mathcal{G}_{crit}$&0.0834 &0.0263 &0.2155 &0.1512\\ \cline{2-7}
& $\hat{\mathcal{G}}_1$ &$\mathcal{G}_1$ &0.1261 & 0.0263 &0.1360  &0.1512 \\ \hline
\multirow{2}{*}{Grid}&$\hat{\mathcal{G}}_1$ & $\mathcal{G}_{crit}$ & 0.0593 &0.0224 &0.0151 &0.0227\\ \cline{2-7} &$\hat{\mathcal{G}}_1$ & $\mathcal{G}_1$ & 0.0593 &0.0224 &0.0151 &0.0227 \\ \hline \hline
\end{tabular}
\caption{Measures of disparity between the DISCRIT graph, the exact
  degree-1 geometric graph, and the critical geometric graph, for the uniform i.i.d.\
  deployment, the randomised lattice deployment, and the deterministic lattice grid.}
\label{tbl:comparison_with_actual_GGs}
\end{table}

Simulation results shown in Table~\ref{tbl:comparison_with_actual_GGs}
show that the DISCRIT output is a very good approximation to the
corresponding critical geometric graph and FNNGG, particularly when
only the interior nodes are considered. Note that the exact degree-1
GG is in fact identical to the critical GG for the i.i.d. and grid
deployments as well as for the interior nodes in the randomised
lattice deployment. This shows that Penrose's
Corollary~\ref{cor:relation between r_1 and r_crit}, though applicable
only asymptotically, practically holds even for finite node densities.


\section{ CGG based Distance  Discretisation} 
\label{sec:distance_discretisation}
In this section we show how ${\cal G}_{crit}$ and DISCRIT can be
useful in obtaining various optimal topologies in a distributed
fashion. Given a graph, the \emph{hop-distance}~\footnote{hop-distance
  is a ``distance'' as it satisfies the properties of non-negativity,
  symmetry, and triangle inequality} between two nodes on the graph is
defined as the minimum number of hops on the graph between the
nodes. We propose to use the hop-distance on ${\cal G}_{crit}$ between
two nodes as a measure of (Euclidean) distance between them. This
technique is distributed since, we already have DISCRIT which is a
distributed construction of ${\cal G}_{crit}$, and hop-distance
calculation can be done using the distributed Bellman-Ford algorithm.


The concept of using hop-distance as a distance measure already exists
in the literature; the important applications being DV-hop routing by
Niculescu and Nath \cite{wsn.niculescu-nath01aps}, and localisation by
Nagpal et al. in \cite{wsn.nagpal-etal03localization}. In all these
methods, the hop-distance is calculated on a specific geometric graph
${\cal G} (\mathbf{V}, R_0)$ where $R_0$ is the communication range of
each node. Note that for a flat terrain and omnidirectional
transmission, the edges in ${\cal G} (\mathbf{V}, R_0)$ join all
possible direct communication neighbours. However we intend to use
hop-distance on ${\cal G}_{crit}$. Our technique is advantageous over
the existing methods like DV-hop in the following ways:
\begin{enumerate}
\item ${\cal G}_{crit}$ is an intrinsic structure of the node layout,
  unlike ${\cal G} (\mathbf{V}, R_0)$ which is dependent on the
  communication parameters. Hence the proposed distance discretisation
  technique is independent of the communication setup.

\item It is shown in \cite{wsn.nagpal-etal03localization} that the
  error in distance estimation is proportional to the radius of the GG
  used for hop-distance calculation. (To see this, consider a line and
  if one has to express distance in integral multiples of $r$, then a
  mean error of $0.5 r$ is obtained.) Thus, since $r_{crit}$ is the
  smallest GG radius which ensures connectivity, using ${\cal
    G}_{crit}$ provides better distance resolution than ${\cal G}
  (\mathbf{V}, R_0)$. Also, unlike a fixed $R_0$, $r_{crit}$ decreases
  with increase in $n$ (see scaling of $r_{crit}$ in
  \cite{wsn.gupta-kumar98connectivity}); thus the distance estimation
  using ${\cal G}_{crit}$ keeps improving with $n$ (also see numerical
  evaluation below).
\end{enumerate}

Given node locations $\mathbf{V}$, let $h_{i,j}$ represent the
hop-distance between $i$ and $j$ on ${\cal G}_{crit}$.  Define
$\rho_{i,j} = \frac{d_{i,j}}{h_{i,j}}$. Note that the proposed
distance discretisation is valid if the hop-distance $h_{i,j}$ is
proportional to the Euclidean distance $d_{i,j}$, i.e., if
$\rho_{i,j}$ is a constant for all $(i,j)$ pairs. Simulation results
show that as $n$ increases, $d_{i,j}$ becomes \emph{roughly}
proportional to $h_{i,j}$.

\vspace{-1mm}
\subsection{Numerical Evaluation}
\begin{figure} [t]
\centering
\begin{minipage}{7.6cm}
\hspace{-5mm}
\includegraphics[width= 8.5cm, height= 4.7cm
]{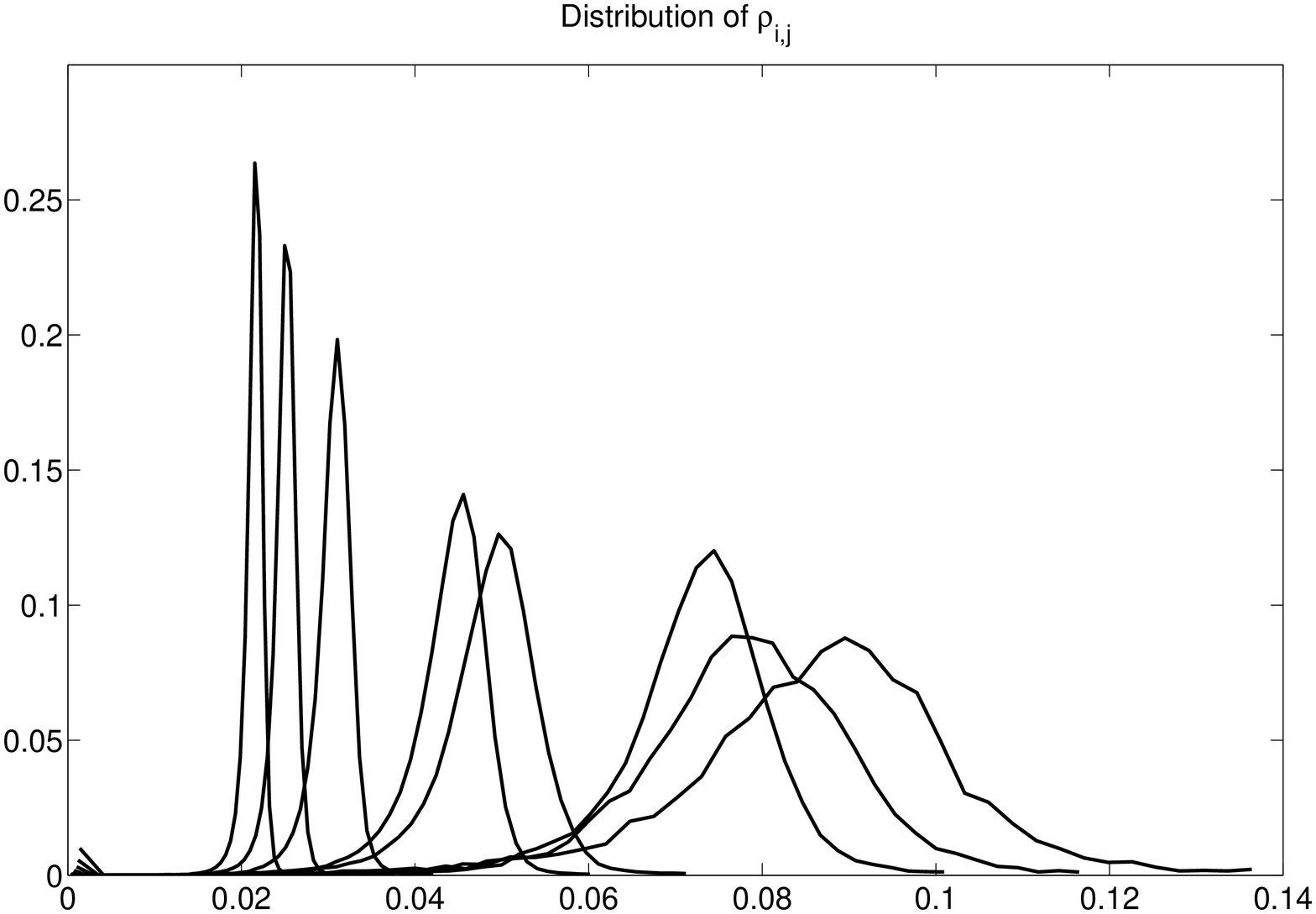}
\caption{Empirical distribution (normalised histogram) of $\rho_{i,j}$
for a sample uniform i.\ i.\ d.\ deployment for each $n$. 8 values of
$n$ are considered between 100 and 5000 nodes}
\label{fig:distribution_rho}
\end{minipage}
\hspace{1mm}
\begin{minipage}{4.7cm}
\includegraphics[width= 5cm, height= 4.5cm
]{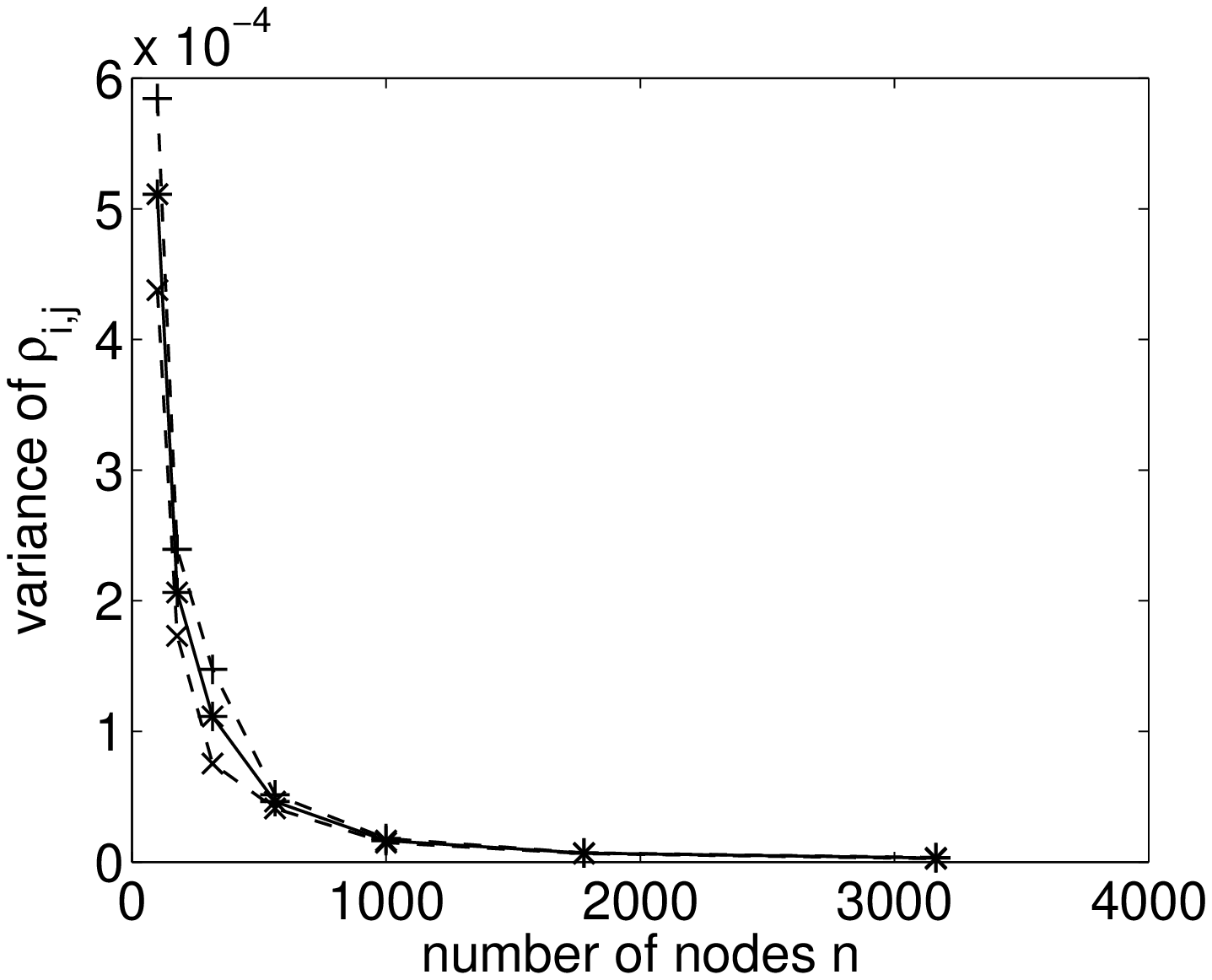}
\caption{Empirical variance of $\rho_{i,j}$. Confidence interval is
  shown for each $n$.}
\label{fig:variance_rho}
\end{minipage}
\hspace{1mm}
\begin{minipage}{4.7cm}
\includegraphics[width= 5cm, height= 4.5cm
]{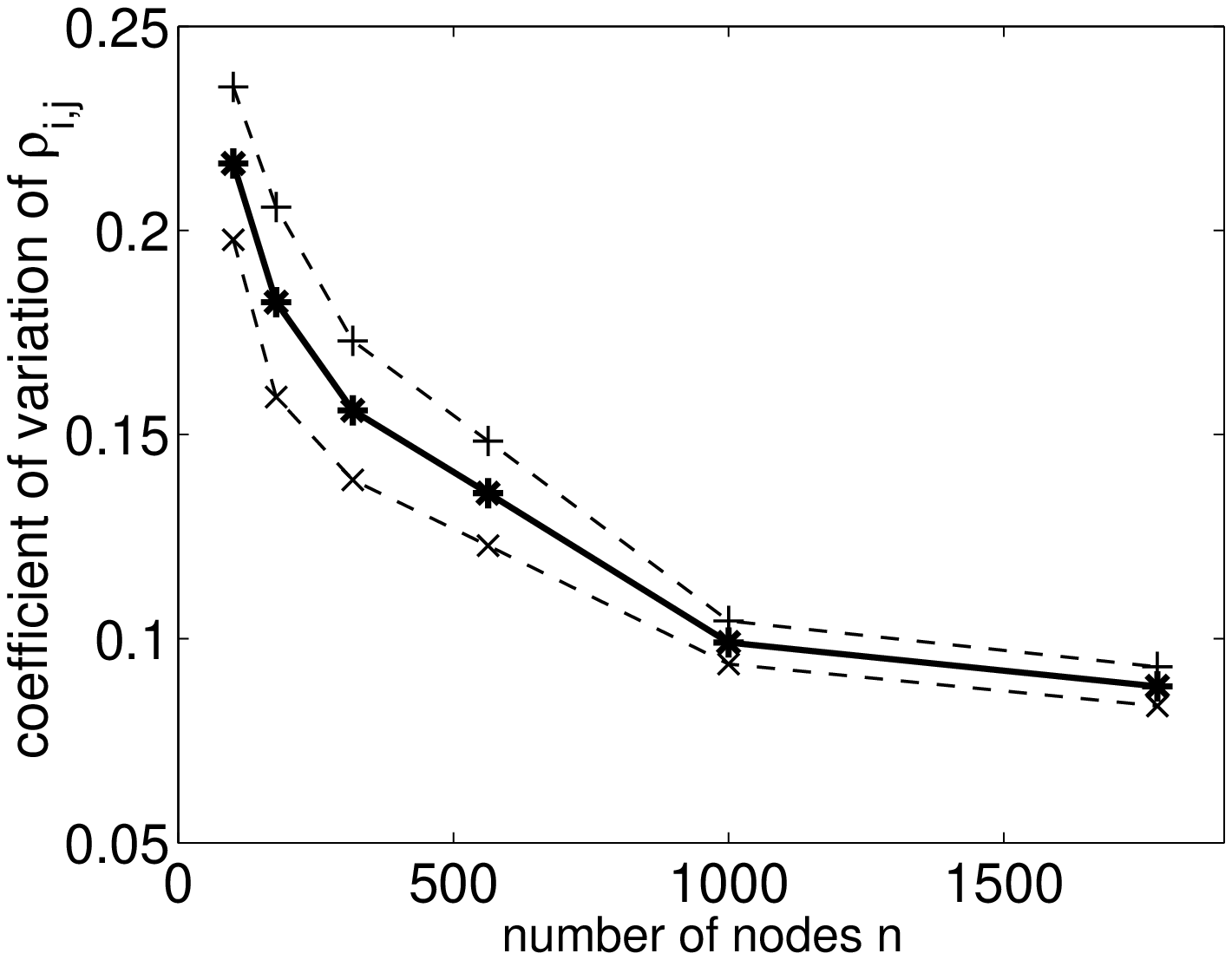}
\caption{Empirical coefficient of variance of $\rho_{i,j}$. Confidence
  interval is shown for each $n$.}
\label{fig:cv_rho}
\end{minipage}
\vspace{-5mm}
\end{figure}

We consider different values of $n$ varying from 100 nodes to 5000
nodes. For each $n$, a sample uniform i.i.d. deployment $\mathbf{V}$
is considered. For each $\mathbf{V}$, the CGG ${\cal
G}_{crit}$ is found, and $\rho_{i,j}$ is evaluated for every node pair
$(i,j)$. Figure~\ref{fig:distribution_rho} shows the empirical
distribution (normalised histogram) of $\rho_{i,j}$ for different
$n$. As $n$ is increased, the support of distribution moves to the
smaller values and the distribution becomes narrower. In other words, the
variation of $\rho_{i,j}$ over node pairs decreases as the node density is increased.

Figure~\ref{fig:variance_rho} and Figure~\ref{fig:cv_rho} show the
plots of empirical variance $\sigma^2_\rho(\mathbf{V})$ and the
empirical \emph{coefficient of variation}~\footnote{The coefficient of
variation is the ratio of standard deviation to the mean.}
$CV_\rho(\mathbf{V})$ of $\rho_{i,j}$ against $n$ (along with the
confidence intervals). The values of $\sigma^2_\rho(\mathbf{V})$ and
$CV_\rho(\mathbf{V})$ are found to decrease as $n$ is
increased. Thus the plots indicate that the $\rho_{i,j}$ becomes
\emph{roughly} a constant for large $n$, thus justifying the proposed distance
discretisation technique.

\vspace{-1mm}
\section{An Example Application: Optimal
  Self-Organisation~\cite{wsn.ramaiyan-etal07optimal-transport-capacity}
} \label{sec:optimal_self_organisation}
\begin{figure} [t]
\centering
\begin{minipage} {8cm}
\centering
\includegraphics[width=8cm, height=4.5cm ]{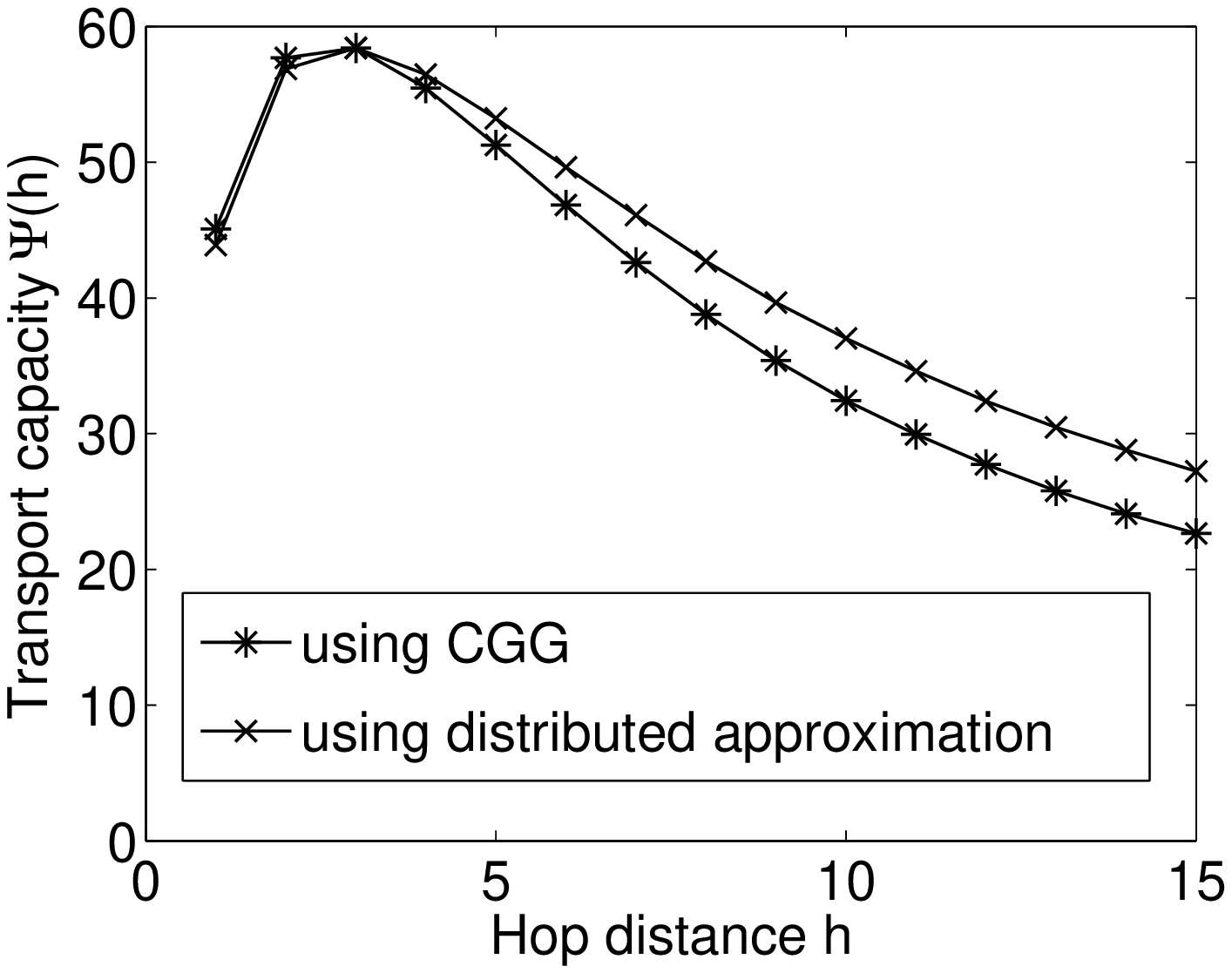}
\caption{Plot of $\Psi_h$ vs. $h$ using centralised ${\cal G}_{crit}$
  and $\hat{{\cal G}}_1$ for hop-distance calculation. The optimal
  hop-distance $h_{opt}=3$ in both cases.}
\label{fig:transport capacity simulation}
\end{minipage}
\hspace{1mm}
\begin{minipage} {8cm}
\centering
\includegraphics[width=8cm, height=4.5cm ]{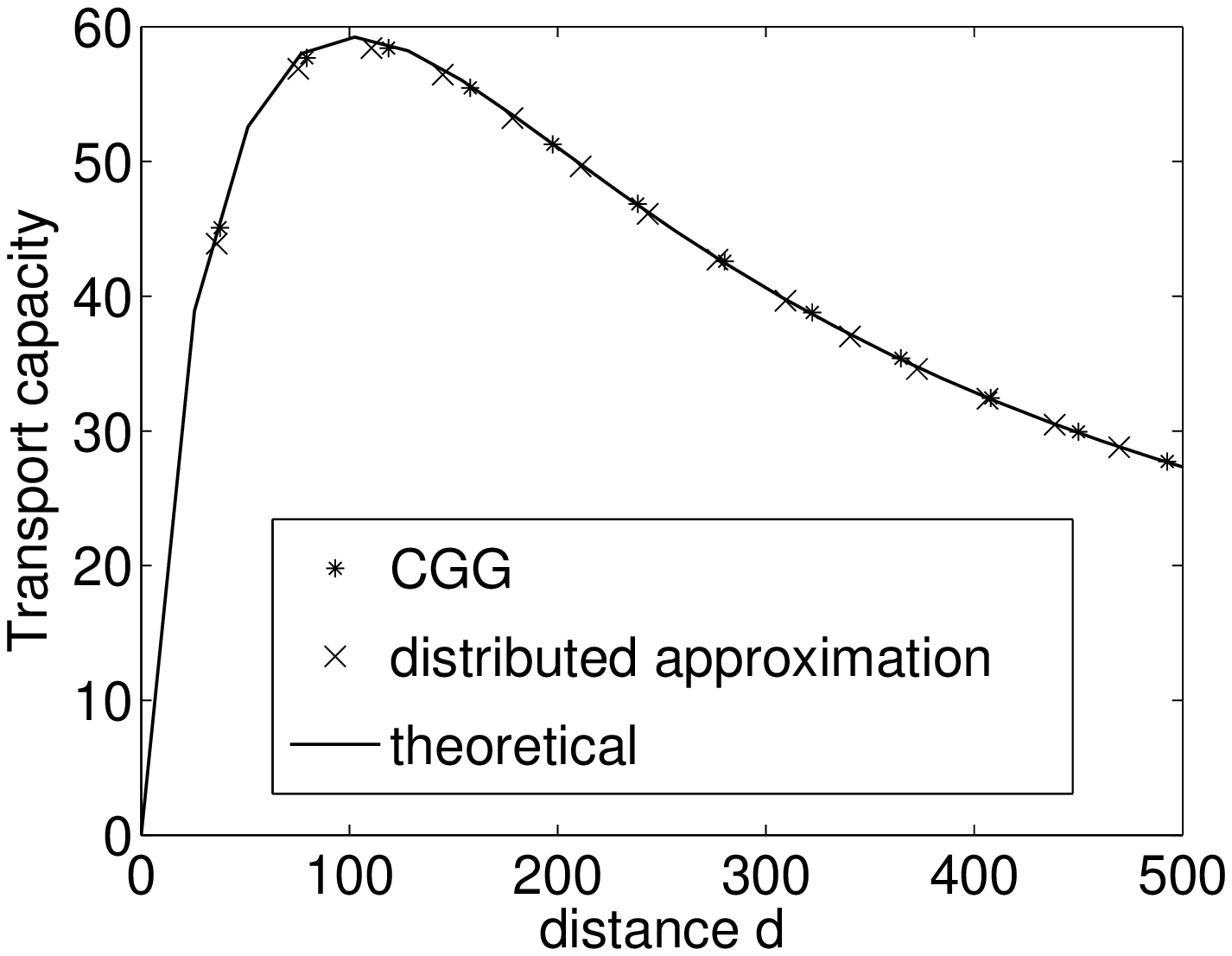}
\caption{Comparison of theoretical Transport capacity with those
  obtained form simulation. The hop-distance $h$ is replaced by mean
  hop-length in $T_h$.}
\label{fig:transport capacity all}
\end{minipage}
\vspace{-5mm}
\end{figure}
Here we illustrate the usefulness of ${\cal G}_{crit}$ based distance
discretisation (and hence of DISCRIT) to self-organisation problems
involving distance information. We consider a self-organisation
problem formulated in
\cite{wsn.ramaiyan-etal07optimal-transport-capacity} of obtaining the
optimal hop-length which maximises the transport capacity on a
\emph{single-cell} dense ad hoc network. There is a dense deployment
of nodes in a limited area. Source-destination pairs are chosen
randomly and the traffic is assumed to be \textit{homogeneous}. A
multihop ad hoc wireless sensor network needs to be self-organised in
such a way that all communication hops are of equal length $d$. The
multiple access protocol is such that only one successful transmission
can occur at any time in the network (i.e., there is \emph{no spatial
  reuse}). Although
\cite{wsn.ramaiyan-etal07optimal-transport-capacity} considers a
fading channel model, we will restrict to the no-fading case. There is
an node transmit power constraint, $P_t$, and the nodes have the
capability of achieving the Shannon capacity over the hop-length $d$,
that is, the bit rate of $W\log \left(1+\frac{\alpha_0 P_t}{d^\eta
    \sigma^2}\right)$ where $W$ is the bandwidth and $\alpha_0$ is a
constant accounting for power gains between the transmitter and
receiver. Under this setup, the aggregate bit rate carried by the
system when all nodes transmit over a distance $d$ per hop, takes the
form (see \cite{wsn.ramaiyan-etal07optimal-transport-capacity})
$\lambda(d) = a \log \left(1+ \frac{\alpha_0 P_t}{d^\eta
    \sigma^2}\right)$ where $a$ is a constant which depends on the
contention parameters. The objective is to maximise the network
\emph{transport capacity} (in bit-meters/sec) given by
\begin{equation} \label{eqn:transport capacity}
\vspace{-2mm}\Psi(d) = d \lambda(d)= a \ d \ \log \left (1+ \frac{ \alpha_0
  P_t}{d^\eta \sigma^2 }\right )\vspace{-1mm}
\end{equation}
over all hop-lengths $d$. It can be seen that there exists an optimal
hop-distance $d_{opt}$ which maximises $\Psi(d)$. The trade-off comes
from the fact that if the network self-organises into short hop
lengths, then the bit rate achieved over a hop is large, but each
packet has to traverse many hops. On the other hand if $d$ is large
then the bit rate over a hop will be small, but fewer hops need to be
traversed.
\vspace{-1mm}
\subsection{A Self-Organisation Heuristic}
We aim to obtain a topology whose hop-lengths are close to
$d_{opt}$. We employ the distance discretisation technique described
in Section~\ref{sec:distance_discretisation} to convert the problem of
finding $d_{opt}$ to one of finding an optimal hop-distance (on ${\cal
  G}_{crit}$), $h_{opt}$, in order to maximise transport capacity. The
resulting self-organisation algorithm is described below.
\begin{enumerate}
\item \textbf{Obtain the critical geometric graph ${\cal G}_{crit}$:} Given the
   node locations, obtain ${\cal G}_{crit}$. This
   is required to obtain the hop-distance information.
\item \textbf{Obtain $h$-hop-distance topology $T_h$:} For $h \ge 1$,
  the $h$-hop-distance topology $T_h$ is obtained by having an edge
  between all node-pairs $(i,j)$ with hop-distance $h_{i,j} =h$ on
  ${\cal G}_{crit}$. The edges (hops) in this topology are considered
  to have $h$ units of distance. $T_1$, thus, denotes the CGG.
\item \textbf{Find the optimal hop-distance $h_{opt}$:} Using each
  topology $T_h$, find the corresponding network transport capacity
  $\Psi_h$. The optimal hop-distance topology $T_{opt}$ (equivalently
  the optimal hop-distance $h_{opt}$) is chosen to be the one which
  maximises $\Psi_h$.
\end{enumerate}
\vspace{-1mm}
\subsection{Numerical Results}
The uniform i.i.d deployment shown in
Figure~\ref{fig:random_comparison_graphs} is considered. We use both
the centralised ${\cal G}_{crit}$ and the distributed $\hat{{\cal
    G}}_1$ for hop-distance calculation. The hop-distances are
calculated using distributed distance-vector routing, using which
$h$-hop topologies $\{T_h\}$ are found.

Nodes are assumed to be saturated and attempt for the channel with a
fixed probability. For each topology $T_h$, the following is
done. Whenever successful, a node transmits randomly to one of its
adjacent nodes in $T_h$, i.e., to a neighbour $h$ hops away on ${\cal
G}_{crit}$. Each node counts the number of bit-meters transmitted by
it. At the end of certain time $t$, the network transport capacity $\Psi_h$
is calculated as the average bit-meters transmitted by all
the nodes in unit time.

The plot of transport capacity $\Psi_h$ against the hop-distance $h$
is shown in Figure~\ref{fig:transport capacity simulation}. The
optimal hop-distance $h_{opt}=3$ when either of ${\cal G}_{crit}$ or
$\hat{{\cal G}}_1$ is used. We may conclude that using $T_3$ could be
optimum for the transport capacity objective. For comparing the result
from the heuristic with the theory, we plot the transport
capacity $\Psi_h$ against the \emph{mean hop-length} in $T_h$, and
then compare it with the theoretical plot obtained using Equation~\ref{eqn:transport capacity}. The plots are shown together in
Figure~\ref{fig:transport capacity all}. The plots being \emph{close}
to each other verify the validity of the proposed heuristic, and the
applicability of the proposed distance discretisation technique.

Thus, for the example scenario above, nodes will use $T_3$ as the communication topology in order to obtain a good transport capacity. That is, each node will communicate directly with nodes 3 hops away in ${\cal G}_{crit}$ (or in DISCRIT output $\hat{\cal G}_1$) as $h_{opt}=3$.

\vspace{-1mm}
\section{Application to Node Localisation} \label{sec:localisation}
\begin{figure*}[t]
\centering
\begin{minipage}{5.5cm}
\includegraphics[width= 5cm, height= 4.5cm
]{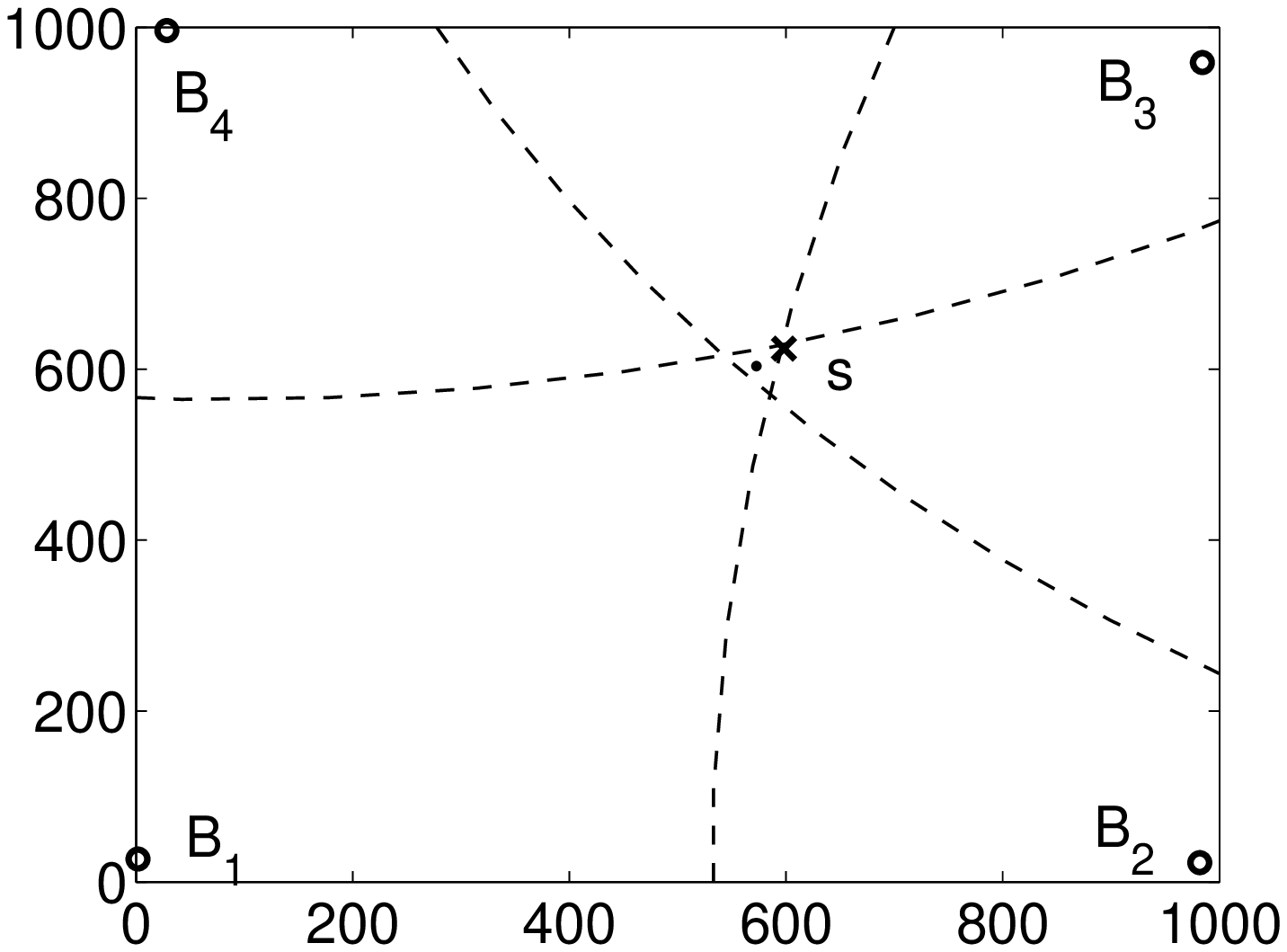}
\caption{Illustration of node localisation for a sample node. The
  actual position is shown as $\times$ and the estimate is shown as
  \textbf{.} in the figure.}
\label{fig:node localisation illustration}
\end{minipage}
\hspace{1mm}
\begin{minipage}{5.5cm}
\includegraphics[width= 5cm, height= 4.5cm
]{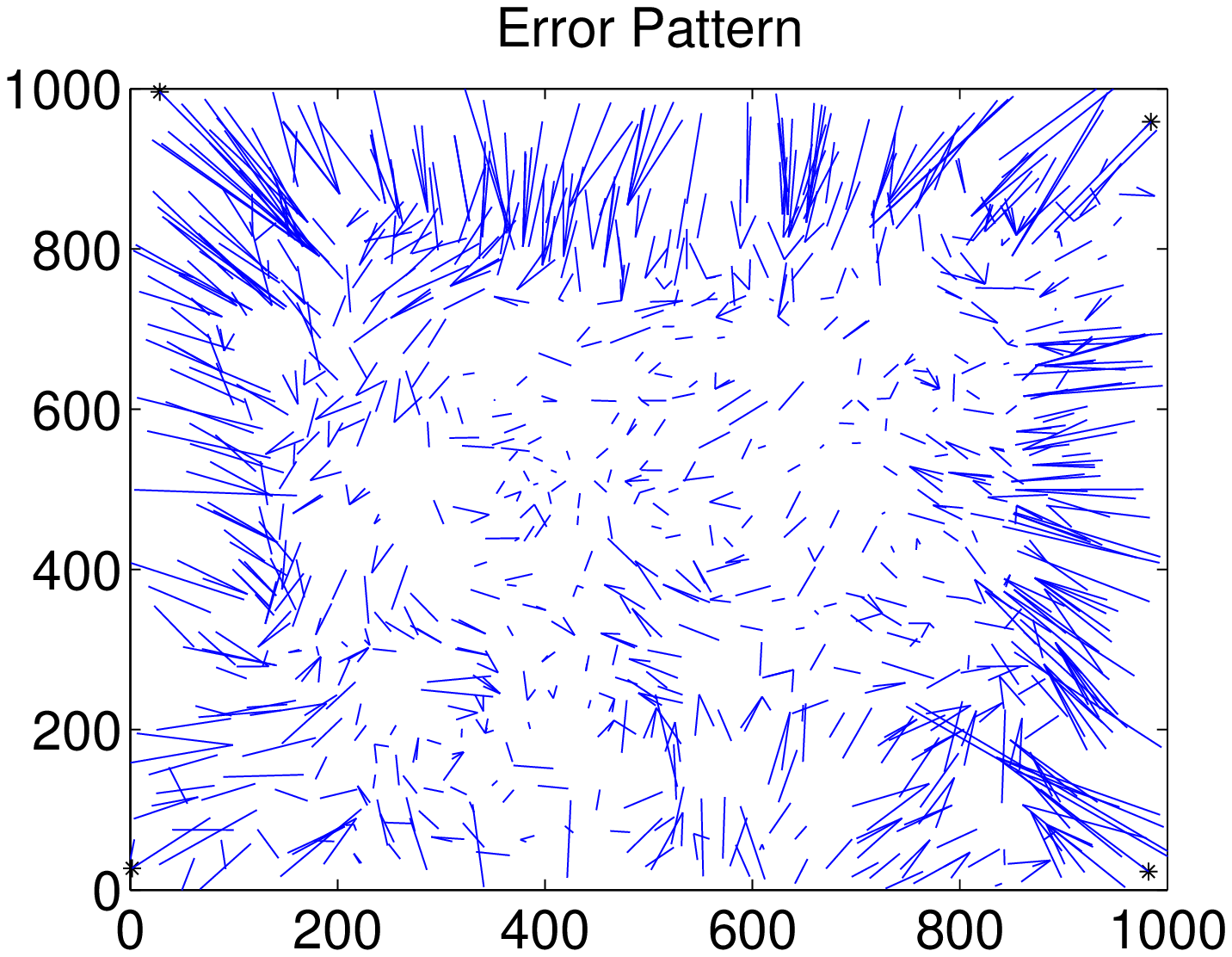}
\caption{Node localisation using exact ${\cal G}_{crit}$: Error pattern.}
\label{fig:error pattern uniform iid}
\end{minipage}
\hspace{1mm}
\begin{minipage}{5.5cm}
\includegraphics[width= 5cm, height= 4.5cm
]{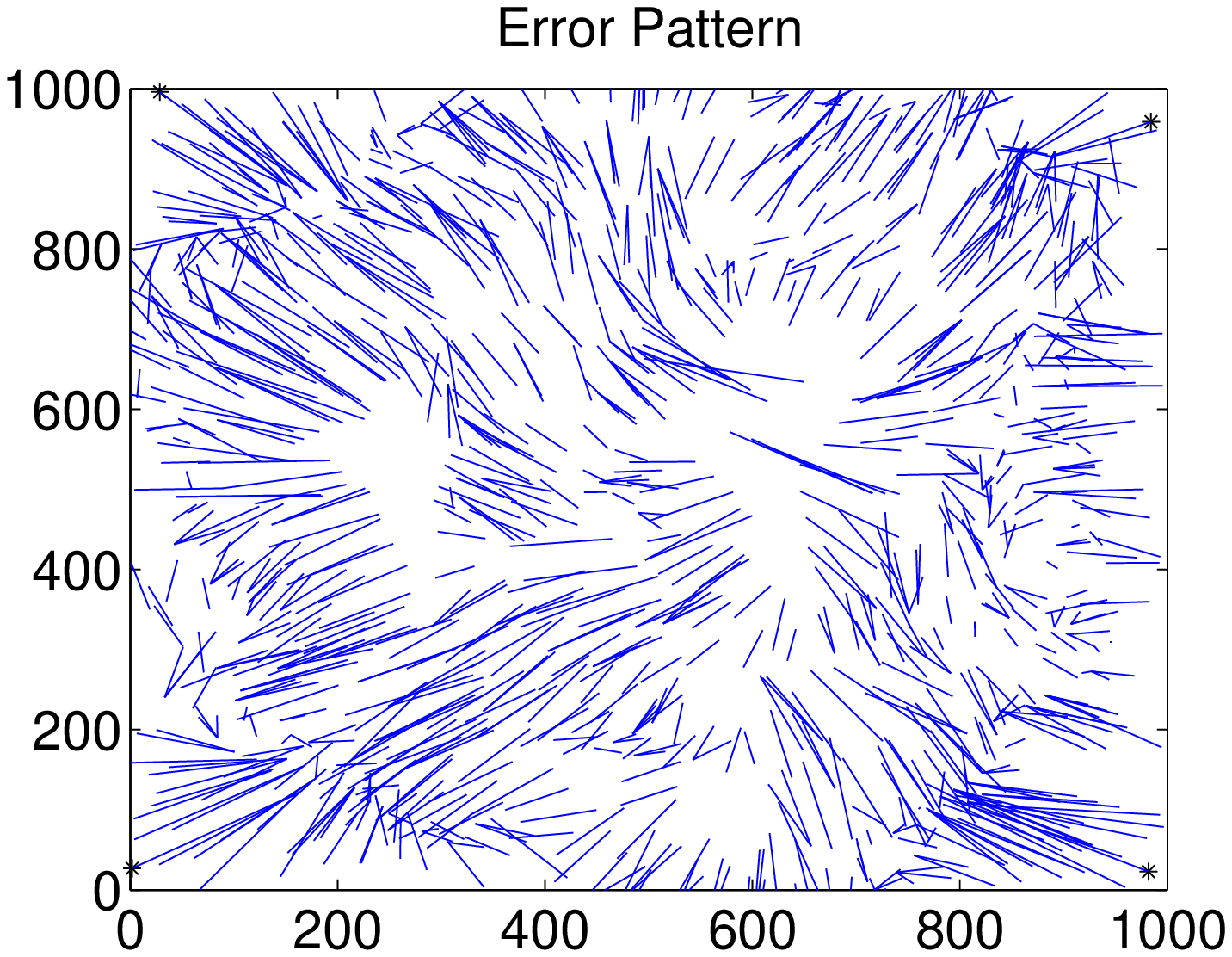}
\caption{Node localisation using distributed $\hat{{\cal G}}_1$: Error pattern.}
\label{fig:error pattern uniform iid distributed}
\end{minipage}
\vspace{-7mm}
\end{figure*}
We illustrate another use of ${\cal G}_{crit}$ and DISCRIT in
estimating the locations of the nodes in a sensor network. The method
described here was conceived independently by us
\cite{wsn.acharya-kumar07self-organisation}, but has been also
reported by Yang et al.\cite{wsn.yang-etal07hcrl}, who call it
\emph{Hop Count Ratio-based Localisation} (HCRL). But the difference
is that, like other existing methods which use hop-distances, HCRL
also uses ${\cal G} (\mathbf{V},R_0)$ (described in
Section~\ref{sec:distance_discretisation}) for hop-distance
calculation, while we use ${\cal G}_{crit}$.
\vspace{-1mm}
\subsection{Theory}
Given the node deployment $\mathbf{V}$, consider $n_B$ beacons (nodes
whose positions are known) given by the set $\{B_1,
B_2,\dots,B_{n_B}\}$. Let $S$ be a node whose location is to be
determined. From the distance discretisation technique described in
Section~\ref{sec:distance_discretisation}, we take $h_{S,B_i} \propto
d_{S,B_i}$ approximately for large $n$. Thus, for any $1\le i,j \le
n_b$, we have \vspace{-1mm}
\begin{equation} \label{eqn:ratio of distances to beacons}
\frac{d_{S,B_i}}{d_{S,B_j}} \approx \frac{h_{S,B_i}}{h_{S,B_j}}=
r_{i,j} \mbox{ (say)}
\end{equation}
Let $(x,y)$, $(x_i,y_i)$, and $(x_j,y_j)$ be the co-ordinates of $S$,
$B_i$, and $B_j$ respectively. Then, taking the approximation in
(\ref{eqn:ratio of distances to beacons}) to be an equality, we have
\vspace{-1mm}
\[\frac{\sqrt{(x-x_i)^2+(y-y_i)^2}}{\sqrt{(x-x_j)^2+(y-y_j)^2}}=r_{i,j} \qquad \mbox {solving which we get}\]
\vspace{-2mm}
\begin{equation} \label{eqn:locus}
\left(1-r_{i,j}^2 \right)x^2 + \left(1-r_{i,j}^2 \right)y^2 -2
\left(x_i-r_{i,j}^2 x_2\right)x -2\left(y_i-r_{i,j}^2 y_j\right)y
+\left(x_i^2 +y_i^2 \right)-r_{i,j}^2\left(x_j^2 +y_j^2\right)=0
\end{equation}
When $r_{i,j}=1$, the equation above is linear and
represents the perpendicular bisector of line joining $B_i$ and
$B_j$, whereas when $r_{i,j} \ne 1$, it represents a circle. In
general, such a circle is called an \emph{Apollonius
circle} \cite{wsn.yang-etal07hcrl}\cite{wsn.durell28apollonius_circle}.
As we need a minimum of 3 such circles to get a point estimate of the
node location, we need at least 4 beacons. Note that 3 Apollonius
circles can be obtained even with 3 beacons, but all 3 intersect at 2
points in general, and hence an additional beacon is required.
\vspace{-1mm}
\subsection{Numerical Results}
We consider the uniform i.i.d.\ deployment of 1000 nodes in a 1~Km
$\times$ 1~Km square shown in
Figure~\ref{fig:random_comparison_graphs}. 4 nodes nearest to 4
corners of the square are taken as beacons (shown as \textbf{o} in
Figure~\ref{fig:node localisation illustration}). A sample node $S$ is
considered for the purpose of illustration (shown as $\times$ in
Figure~\ref{fig:node localisation illustration}). The hop-distance
calculation is made on the corresponding ${\cal G}_{crit}$. 3 of the 6
possible Apollonius circles are shown in Figure~\ref{fig:node
  localisation illustration}. The circles correspond to beacon-pairs
$(1,2)$, $(1,3)$ and $(1,4)$. (Note that if $h_{i,j}$ were exactly
proportional to $d_{i,j}$, then all the Apollonius circles would
intersect at a point, which would coincide with the actual location
$\times$.) To obtain an estimate, we use an inbuilt MATLAB function
which solves an optimisation problem involving the circle
equations. The estimate is shown as \textbf{.} in Figure~\ref{fig:node
  localisation illustration}. The \emph{error} in the position
estimation, i.e., the distance between the node's actual position and
the estimate is 41~m (comparable with $r_{crit}$=56.7~m).

To observe the localisation performance for all nodes, an error
pattern is obtained by joining each node's actual position and its
estimate. The error pattern is shown in Figure~\ref{fig:error pattern
  uniform iid}. The mean estimation error is 55~m. The same experiment
is repeated by performing hop-distance calculation on the DISCRIT
output $\hat{{\cal G}}_1$. The resulting error pattern is shown in
Figure~\ref{fig:error pattern uniform iid distributed}. The mean error
is 74~m in this case, which is higher compared to that of centralised
${\cal G}_{crit}$. In both the cases, the error is large towards the
edges because of smaller node density at the edges and since the
distance discretisation technique is valid for large node densities
(see Section~\ref{sec:distance_discretisation}).

\vspace{-1mm}
\section{Conclusion} \label{sec:conclusion} The paper provides
DISCRIT: a distributed algorithm to approximate ${\cal G}_{crit}$,
using link weights obtained from the Hello-protocol-based neighbour
discovery. The validity of DISCRIT is shown (with high probability)
for a dense uniform i.\ i.\ d.\ deployment, by making use of Penrose's
result and the spatial homogeneity of the deployment. Simulation
results are shown for other types of deployments too, which indicate
that the algorithm provides  good approximations of ${\cal
  G}_{crit}$. Using ${\cal G}_{crit}$ (and DISCRIT), a distributed
technique of associating distances to links joining node pairs is also
proposed. Example applications of ${\cal G}_{crit}$ based distance
discretisation are shown for a self-organisation problem of obtaining
the optimal hop-length which maximises transport capacity for a dense
ad hoc network operated as a single-cell, and also for  node
localisation.

In related work, we have also considered the problem of anisotropic
antenna radiation patterns, which renders invalid the direct use of
Hello counts as described in this paper. However, assuming that
antenna patterns are randomly and uniformly oriented, nodes can
locally cooperate to address the anisotropy problem. 

Other extensions of this work could include handling spatially
non-homogeneous deployments, which in turn will also help reduce the
\emph{edge effect} seen in DISCRIT outputs. Other practical issues to
be examined are self-tuning of parameters such as the Hello
transmission latency and node transmit power for improved performance
of DISCRIT.

The algorithms and techniques proposed in the paper, apart from being
distributed and asynchronous, require only limited capabilities from
the nodes. The nodes are not equipped with position finding devices
such as GPS. Receive Signal Strength (RSS) based techniques for
distance estimation, which require accurate power measurement and
scheduling, are also not used. Thus, our approach can provide a simple
yet effective way of self-organisation in sensor networks.
\vspace{-1mm} \bibliographystyle{IEEEtran}
\bibliography{wsn.self-organization}

\newpage
\appendix

\begin{appendices}
%
\section{Proof of Theorem~\ref{thm:algorithm}} \label{appendix:proof}
\textbf{Theorem:}
If ${\cal G}_1$ is connected, then the algorithm converges to ${\cal G}_1$, i.e.,
  ${\cal G}_A = {\cal G}_1$, in at most $D$ iterations, where $D$ is
  the \emph{hop diameter} of ${\cal G}_1$.
\begin{IEEEproof}
\renewcommand{\thelemma}{\thesection.\arabic{lemma}}
The proof proceeds in a few lemmas.
\begin{lemma} \label{lemma 1}
For all $k$, for all $i \in N$, $r^{(k)}(i) = r^{(0)}(j)$ for some $j
\in N$ ($j$ depends on $i$ and $k$).
\end{lemma}
\begin{IEEEproof}
\emph{Induction on k}: From \textbf{Step 3} of the algorithm, we have
 $r^{(k+1)}(i)= \max\{r^{(k)}(j): j\in S^{(k)}(i)\}$. Therefore,
 \[r^{(k)}(i) = r^{(k-1)}(i_1)= \dots = r^{(1)}(i_{k-1}) =r^{(0)}(j)\]
 for some $i_1,i_2, \dots, j \in N$.
\end{IEEEproof}
\begin{lemma}  \label{lemma 2}
For all $i \in N $, for all $k$, $r^{(k)}(i) \leq r_1$.
\end{lemma}
\begin{IEEEproof}
From Lemma~\ref{lemma 1}, $r^{(k)}(i) = r^{(0)}(j)$ for some $j \in
N$. Therefore
\[r^{(k)}(i)= r^{(0)}(j)= \min_{k \in N, k\neq j}
\{d_{j,k}\} \leq \max_{j\in N}\{\min_{k \in N, k\neq j} \{d_{j,k}\}\}
= r_1\]
\end{IEEEproof}
\begin{lemma} \label{lemma 3}
If $r^{(k)}(i) = r_1$ for some $k$, then for $s=1,2,3,\dots$,
$r^{(k+s)}(i)=r_1$
\end{lemma}
\begin{IEEEproof}
Note that $i \in N^{(k)} (i)$ since $d_{i,i} =0$, and hence $i \in
S^{(k)}(i)$. Thus
\[r^{(k+1)}(i)= \max\{r^{(k)}(j):j \in S^{(k)}(i)\} \geq r^{(k)}(i) = r_1\]
But from Lemma~\ref{lemma 2}, $r^{(k+1)}(i)
\leq r_1$, therefore $r^{(k+1)}(i) = r_1$.

The argument can be
extended to show $r^{(k+s)}(i) = r_1$ for $s=2,3,\dots$.
\end{IEEEproof}

Let $l =\arg \max_{i \in N}\{r^{(0)}(i)\}$. Then, $r^{(0)}(l)= r_1$. Let
$C(h) = \{i:i$ is connected to $l$ in $h$ hops in ${\cal G}_1\}$.

\begin{lemma} \label{lemma 4}
 For all $i \in C(h)$, $r^{(h)}(i)=r_1$.
\end{lemma}
\begin{IEEEproof}
\emph{Induction on $h$:} $i \in C(1)$ means that $d_{i,l}\leq r_1 =
r^{(0)}(l)$. Therefore, $i \in N^{0}(l) \Rightarrow l \in S^{(0)}
(i)$, and hence, $r^{(1)}(i) \geq r^{(0)}(l) =r_1$. But from
Lemma~\ref{lemma 2}, $r^{(1)}(i) \leq r_1$. Therefore, $r^{(1)}(i) =
r_1$.

Assume, for all $i \in C(h)$, $r^{(h)}(i) =r_1$. Now, for any $i
\in C(h+1)$, there exists a $j \in C(h)$ such that $d_{i,j} \leq
r_1$. Thus $r^{(h)}(j)=r_1 \Rightarrow i \in N^{(h)}(j)$, and hence $j \in
S^{(h)} (i)$. Therefore $r^{(h+1)}(i) =\max\{r^{(h)}(t): t \in S^{(h)}
(i)\}\geq r^{(h)}(j) = r_1$. Again from Lemma~\ref{lemma 2},
$r^{(h+1)}(i)= r_1$ for all $i \in C(h+1)$.

Lemma~\ref{lemma 4} now
follows from the induction principle.
\end{IEEEproof}
\begin{lemma}  \label{lemma 5}
Suppose ${\cal G}_1$ is \emph{connected}. Let $m$ be the maximum
number of hops in the least-hop path on ${\cal G}_1$ between any node $i$ and $l$ ($m$
is the \emph{hop radius} of ${\cal G}_1$ centered at $l$). Then the
algorithm converges in $m$ iterations. Moreover $r(i)=r_1$ for all
$i$.
\end{lemma}
\begin{IEEEproof}
Take any node $i \in N$. Since ${\cal G}_1$ is
connected, $i \in C(h)$ for some $h \leq m$. Now from
Lemma~\ref{lemma 4}, $r^{(h)}(i)=r_1$ and therefore from
Lemma~\ref{lemma 3}, $r^{(m)}(i)=r_1$ as $m \geq h$.

Now at $m^{th}$ iteration, we have
$r^{(m)}(i)=r_1$ for all $i \in N$. Therefore from Lemma~\ref{lemma 2},
$r^{(m+1)}(i)=r_1$ for all $i$.

Thus at \textbf{Step 4}, the algorithm terminates and $r(i)=r_1$ for
all $i \in N$.
\end{IEEEproof}

Thus, from Lemma~\ref{lemma 5}, $r(i)= r_1$, and $N(i) = \{j: d_{i,j}
\leq r_1\}$. Now from \textbf{Step 5} of the algorithm, it can be seen
that ${\cal G}_A = {\cal G}_1$. Theorem~\ref{thm:algorithm} follows by
observing the fact that $m \leq D$, the hop diameter of ${\cal
G}_1$.
\end{IEEEproof}
%
\section{Proof of Theorem~\ref{thm:homogeneity}} \label{appendix:proof of homogeneity}
\textbf{Theorem:}
For any $\epsilon >0$ however small,
\[\lim_{n \to \infty} {\cal P}^n \left\{ \mathbf{V}: \frac{n}{\mid {\cal A} \mid} (1-\epsilon) \le \frac{N_r(x;\mathbf{V})}{\pi r^2} \le \frac{n}{\mid {\cal A} \mid} (1+\epsilon) \mbox{ for every } x \in \tilde{\cal A}(r) \right\} = 1\]
\begin{IEEEproof}
  Note that the expression contained within ${\cal P}^n\{.\}$ involves
  an intersection of (uncountably) infinite events, where each event
  corresponds to a bound on $N_r(x;\mathbf{V})$ for each point $x \in
  \tilde{\cal A}(r)$. For this purpose, we make use of uniform
  convergence of weak law of large numbers, given by
  \emph{Vapnik-Chervonenkis} theorem.

Given a set $U$, let ${\cal C}$ be a collection of subsets of $U$. Let
$F$ be a finite subset of $U$.  $F$ is said to be \emph{shattered} by
${\cal C}$ if for every subset $G$ of $F$, there exists a subset $C \in
{\cal C}$ such that $F \cap C = G$. The \emph{VC-dimension} of ${\cal C}$
is defined as the supremum of sizes of all finite sets ($F$s) that can be
shattered by ${\cal C}$.

\begin{theorem} [Vapnik and Chervonenkis] \label{thm:vapnik}
If ${\cal C}$ is a set of finite VC-dimension $d_{\cal C}$, and $\{X_i\}_{i=1}^m$ be a
sequence of i.i.d random variables taking values in $U$ with common probability
distribution $P$, then for every $\epsilon', \delta >0$
\[ \prob{\sup_{C \in {\cal C}} \left| \frac{\sum_{i=1}^m I_{\{X_i \in
    C\}}}{m} - P(C) \right|  \le \epsilon'}  \ge 1- \delta \]
whenever
\[m \ge \max\{\frac{d_{\cal C}}{\epsilon'}\log \frac{16 e}{\epsilon'},
\frac{4}{\epsilon'}\log \frac{2}{\delta} \}\] $\hfill \qed$
\end{theorem}

Here $I_{\{.\}}$ represents the indicator function. For our purpose,
let $U$ represent $\mathbf{R}^2$. Then $F$ represents a finite set of
points in $\mathbf{R}^2$. Define a \emph{closed disc} of radius $r>0$
and centre $x$ as $D_r(x):= \{y \in \mathbf{R}^2: \parallel x
-y \parallel \le r\}$. Let ${\cal C}$ represent a collection of all
closed discs in $\mathbf{R}^2$, i.e., ${\cal C}:=\{D_r(x): x \in
\mathbf{R}^2, r>0\}$.  Gupta and Kumar in
\cite{wsn.gupta-kumar00capacity} have shown the following.

\begin{lemma} \label{lemma:vc-dimension}
The VC-dimension of set of all closed discs in $\mathbf{R}^2$ is 3, i.e., $d_{\cal C}$ =3. $\hfill \qed$
\end{lemma}

By definition $D_r(x) \in {\cal C}$ for every $x \in \tilde{\cal
  A}(r)$. Now if we take the common probability distribution $P$ in
Theorem~\ref{thm:vapnik} to be the uniform measure ${\cal P}$ on
${\cal A}$, then it can be seen that i.i.d. random variable sequence
$\{X_i\}$ is nothing but a uniform i.i.d. deployment ${\mathbf{V}}$ on
${\cal A}$. Now observing that ${\cal P}\{D_r(x)\} = \frac{\pi
  r^2}{\mid {\cal A} \mid}$ for all $x \in \tilde{\cal A}(r)$,
Theorem~\ref{thm:vapnik} and Lemma~\ref{lemma:vc-dimension} give rise
to the following result.
\begin{corollary} \label{cor:vapnik}
For every $\epsilon', \delta >0$
\[ {\cal P}^n \left \{  \left| \frac{\sum_{i=1}^n I_{\{V_i \in
    D_r(x)\}}}{n} - \frac{\pi r^2}{\mid {\cal A}\mid} \right|  \le \epsilon'  \qquad \mbox{for every $x \in \tilde{\cal A}(r)$}\right \}  \ge 1- \delta \]
\[\mbox{whenever \qquad} n \ge \max\{\frac{3}{\epsilon'}\log \frac{16 e}{\epsilon'},
\frac{4}{\epsilon'}\log \frac{2}{\delta} \}\] $\hfill \qed$
\end{corollary}
\begin{eqnarray*}
\mbox{Now} \qquad \left| \frac{\sum_{i=1}^n I_{\{V_i \in D_r(x)\}}}{n} - \frac{\pi r^2}{|{\cal A}|} \right| \le \epsilon' &\Leftrightarrow&
    n(\frac{\pi r^2}{|{\cal A}|} - \epsilon') \le \sum_{i=1}^n I_{\{V_i \in D_r(x)\}} \le n(\frac{\pi r^2}{|{\cal A}|}
    +\epsilon')\\
    &\Leftrightarrow&  \frac{n}{|{\cal A}|} \left(1- \frac{|{\cal A}| \epsilon'}{\pi r^2} \right) \le \frac{\sum_{i=1}^n I_{\{V_i \in D_r(x)\}}}{\pi r^2}\le \frac{n}{|{\cal A}|} \left(1+ \frac{|{\cal A}| \epsilon'}{\pi r^2} \right)
\end{eqnarray*}
Thus Corollary~\ref{cor:vapnik} becomes \\
For every $\epsilon', \delta >0$
\[ {\cal P}^n \left \{  \frac{n}{|{\cal A}|} \left(1- \frac{|{\cal A}| \epsilon'}{\pi r^2} \right) \le \frac{\sum_{i=1}^n I_{\{V_i \in D_r(x)\}}}{\pi r^2}\le \frac{n}{|{\cal A}|} \left(1+ \frac{|{\cal A}| \epsilon'}{\pi r^2} \right)  \qquad \mbox{for every $x \in \tilde{\cal A}(r)$}\right \}  \ge 1- \delta \]
\[\mbox{whenever \qquad} n \ge \max\{\frac{3}{\epsilon'}\log \frac{16 e}{\epsilon'},
\frac{4}{\epsilon'}\log \frac{2}{\delta} \}\]
Theorem~\ref{thm:homogeneity} now follows by taking $\epsilon: = \frac{|{\cal A}|\epsilon'}{\pi r^2}$, and observing that $N_r(x;\mathbf{V})=\sum_{i=1}^n I_{\{V_i \in D_r(x)\}}$, the number of nodes in $D_r(x)$.
\end{IEEEproof}
\end{appendices}

\end{document}